\documentclass[final,3p,times,onecolumn]{elsarticle}

\usepackage{graphicx}
\usepackage{subcaption}
\usepackage{amsmath, amssymb, amsfonts, amsthm}
\usepackage{siunitx}  
\usepackage{booktabs} 
\usepackage{hyperref} 
\usepackage{xurl} 
\usepackage{enumitem} 
\usepackage[final]{microtype}
\usepackage{tikz}
\usepackage{pgfplots}
\usepackage{xcolor}
\usepgfplotslibrary{groupplots}
\usetikzlibrary{calc}
\pgfplotsset{compat=1.18}

\AtBeginEnvironment{thebibliography}{\sloppy}

\newcounter{criterion}
\renewcommand{\thecriterion}{C\arabic{criterion}}

\newcommand{\name}{BERS}
\newcommand{\cmark}{\checkmark}
\newcommand{\rotatext}[1]{\rotatebox[origin=c]{90}{#1}}
\newif\ifshowrafacomments
\showrafacommentsfalse

\newenvironment{DIFnomarkup}{}{}
\begin{document}
\begin{frontmatter}
\title{BERS: A Continuous Hybrid Evolutionary--Variational Algorithm for Maritime Weather Routing with Just-in-Time Arrival}

\author[ie]{Daniel Precioso\corref{corauthor}}
\cortext[corauthor]{Corresponding author}
\ead{daniel.precioso@ie.edu}  

\author[ie]{Francisco Suárez}
\ead{francisco.suarez@ie.edu}  

\author[uca]{Javier Jiménez de la Jara}
\ead{javier.jimenezdelajara@uca.es}  

\author[ie]{Rafael Ballester-Ripoll}
\ead{rballester@faculty.ie.edu}

\author[ie]{David Gómez-Ullate}
\ead{dgomezullate@faculty.ie.edu}

\affiliation[ie]{
    organization={School of Science and Technology, IE University},
    addressline={Paseo de la Castellana, 259},
    postcode={28046},
    city={Madrid},
    country={Spain}
    }
\affiliation[uca]{
    organization={Department of Computer Science, Higher School of Engineering, Universidad de Cádiz},
    addressline={Av. Universidad de Cádiz, 10},
    postcode={11519},
    city={Puerto Real, Cádiz},
    country={Spain}
    }

\begin{abstract}
\begin{sloppypar}
Maritime weather routing must optimize route geometry under dynamic wind-wave conditions, obstacle constraints, and fixed-arrival requirements. We present B\'ezier Evolve and Refine Strategy (\name{}), a two-stage framework that combines global evolutionary search (CMA-ES) with local variational refinement (FMS). Routes are parametrized as B\'ezier curves and evaluated with dense along-path sampling, which resolves mid-segment effects. On synthetic benchmarks that stress seven operational criteria, \name{} matches or improves published baselines under challenging flow fields and land geometries. We then benchmark the method on real-ocean scenarios from the SWOPP3 test bed, using hourly ERA5 for 366 daily departures in 2024 across two trans-oceanic corridors, with a publicly available analytical performance model of an 88~m cargo vessel with rigid wingsails. Mean propulsive energy falls by 23--59\% relative to great-circle routes, reaching 75\% when route optimization and wingsails are combined. FMS improves on CMA-ES in at least 95\% of departures in every configuration, demonstrating systematic local improvement. On the SWOPP3 benchmark, BERS ranked within 1.1\% of the best-performing submission and reported the lowest computation time by more than one order of magnitude. \name{} offers a practical and scalable basis for just-in-time, energy-efficient weather routing.
\end{sloppypar}
\end{abstract}


\end{frontmatter}


\section{Introduction}
\label{sec:introduction}

Weather routing is a challenging optimization problem that focuses on determining efficient maritime routes while taking into account environmental factors such as wind, waves, currents, and other dynamic meteorological conditions. Unlike traditional shortest-path problems, which operate on static networks, weather routing is complicated by continuous changes in input data and the need to balance multiple objectives \cite{Precioso2023Thesis}. This problem is of significant practical importance in modern shipping, where optimizing routes can reduce fuel consumption, travel time, and emissions, while ensuring navigational safety. A key aspect of this problem is to find solutions that balance competing goals, such as minimizing travel time and fuel consumption, often within large and uncertain search spaces \cite{szlapczynski2023ship}. Advances in algorithm design and computational methods have made significant progress in the field, with meta-heuristics \cite{Grandcolas2022}, techniques to incorporate user preferences \cite{Jimenez2024, mannarini2024visir}, and evolutionary approaches \cite{Precioso2024, Zhao2022}.

Weather routing has become a key operational measure for improving both the economic and environmental performance of maritime transport. By continuously optimizing the course and speed of the vessel based on forecasts of wind, waves, and currents, routing systems can reduce fuel consumption and emissions without requiring hardware modifications. According to the International Maritime Organization (IMO) \cite{imo_greenvoyage2050_weatherrouting}, the expected efficiency improvement from weather routing typically ranges between 0 and 5\%, depending on the type of vessel and the characteristics of the route. More recent studies report average savings of 4--7\% when routing is coupled with voyage speed and ETA optimization \cite{sofar2024}. Weather routing companies report average fuel savings of 5.5\%, which is roughly \$ 17{,}700 per voyage for deep-sea bulk carriers \cite{sofar2024}. These figures refer mainly to conventional cargo vessels on established routes. Larger savings are possible for vessels with wind-assisted propulsion, because the route then also determines how much thrust the sails can deliver; the savings reported in Section~\ref{sec:results-real} should be read in that context.

These reductions in fuel translate directly into lower greenhouse-gas emissions: each tonne of avoided marine fuel prevents approximately 3.114 tonnes of CO$_2$ from being emitted \cite{imo_ghg_factors, imo_2023_ghg_strategy}. Therefore, a 5\% fuel reduction on a typical 20-day ocean voyage consuming 600~t of fuel would avoid about 94~t of CO$_2$. The combination of economic incentives and regulatory pressure under initiatives such as the IMO GHG strategy reinforces the relevance of accurate and adaptive routing algorithms as a decarbonization lever in maritime operations.

Taken together, these demonstrated operational gains and decarbonization pressures underscore a central methodological challenge: developing a routing algorithm that can simultaneously meet several technical and operational requirements. To guide our work, we identified seven key design goals that a comprehensive weather routing algorithm should ideally fulfil:

\begin{itemize}[label={\thecriterion}]
    \refstepcounter{criterion}\item \label{cr:continuous-space} \textbf{Continuous in Space:} The algorithm must resolve environmental and constraint effects along each route segment with sufficient spatial resolution, so relevant mid-segment phenomena are not skipped. This concerns not only how trajectories are represented (grid-based or not), but also how densely conditions are evaluated between waypoints.
    \refstepcounter{criterion}\item \label{cr:obstacle-avoidance} \textbf{Obstacle Avoidance:} Safe navigation must be ensured by avoiding landmasses, restricted zones, and other hazards.
    \refstepcounter{criterion}\item \label{cr:dynamic-conditions} \textbf{Dynamic Conditions:} The algorithm should adapt to changing environmental factors, such as evolving wind patterns, ocean currents, and waves.
    \refstepcounter{criterion}\item \label{cr:flexible-cost} \textbf{Flexible Cost Functions:} It should accommodate various optimization goals, such as minimizing travel time, fuel consumption, safety, or a combination of them.
    \refstepcounter{criterion}\item \label{cr:constant-power} \textbf{Constant Engine Load:} The method must handle constant engine power operations, which reflect real ship performance and typically focus on minimizing travel time.
    \refstepcounter{criterion}\item \label{cr:jit} \textbf{Just-in-Time Arrival:} The algorithm should allow adherence to predetermined travel durations, allowing adjustments to optimize fuel savings while ensuring timely arrival.
    \refstepcounter{criterion}\item \label{cr:local-opt} \textbf{Local Optimality:} The approach should find the best possible solution within its neighbourhood.
\end{itemize}

\noindent These criteria are design objectives against which the framework is evaluated, rather than guarantees that every final route satisfies in every tested configuration. Table~\ref{tab:criteria} provides a centralized summary and the sections where each criterion is assessed.

\begin{table}[t]
\centering
\begin{tabular}{clc}
\toprule
Criterion & Description & Assessed in \\
\midrule
\ref{cr:continuous-space} & Continuous spatial resolution via dense sampling & \S\ref{sec:bezier}, \S\ref{sec:results-literature}, \S\ref{sec:results-land} \\
\ref{cr:obstacle-avoidance} & Land and obstacle avoidance & \S\ref{sec:land-avoidance}, \S\ref{sec:results-land}, \S\ref{sec:rw-routes} \\
\ref{cr:dynamic-conditions} & Adaptation to dynamic wind/wave conditions & \S\ref{sec:cost-function}, \S\ref{sec:results-literature}, \S\ref{sec:rw-seasonal} \\
\ref{cr:flexible-cost} & Flexible cost functions (time, fuel, black-box) & \S\ref{sec:cost-function}, \S\ref{sec:results-literature}, \S\ref{sec:results-real} \\
\ref{cr:constant-power} & Constant engine load operation & \S\ref{sec:cost-function-time}, \S\ref{sec:ablation} \\
\ref{cr:jit} & Just-in-time arrival compliance & \S\ref{sec:cost-function}, \S\ref{sec:results-literature}, \S\ref{sec:rw-setup} \\
\ref{cr:local-opt} & Local optimality pursued through variational refinement & \S\ref{sec:algorithm-fms}, \S\ref{sec:fms-second-order}, \S\ref{sec:rw-local-opt} \\
\bottomrule
\end{tabular}
\caption{Summary of the seven design criteria for maritime weather routing algorithms and the sections where each is assessed.}
\label{tab:criteria}
\end{table}

Existing algorithms tend to specialize in meeting some of these criteria but often struggle to address others, resulting in trade-offs that limit their applicability. We identify three broad families of methods used in weather routing: variational optimization, graph-based and dynamic-programming methods, and evolutionary algorithms.

\begin{table*}[htbp]
\centering
\begin{tabular}{llccccccc}
\toprule
\textbf{Reference} & \textbf{Algorithm} & \rotatext{Continuous in Space} & \rotatext{Obstacle Avoidance} & \rotatext{Dynamic Conditions} & \rotatext{Flexible Cost Function} & \rotatext{Constant Engine Load} & \rotatext{Just-in-Time Arrival} & \rotatext{Local Optimality} \\
\midrule
FMS (DNJ) \cite{Ferraro2021}            & Variational                & \cmark  &        &        &        & \cmark & \cmark  &  (\cmark) \\ \midrule
VISIR-2 \cite{mannarini2024visir} & Graph Search               &         & \cmark & \cmark & \cmark &        &  & \cmark$^\star$  \\
PRM \cite{Charalambopoulos2023}   & Graph Search               &         & \cmark & \cmark & \cmark & \cmark &    & \cmark$^\star$ \\
SIMROUTE \cite{Grifoll2022}       & Graph Search               &         & \cmark & \cmark & \cmark & \cmark &    & \cmark$^\star$ \\ \midrule
HADAD \cite{Jimenez2024}          & Graph + Var.        & \cmark & \cmark & \cmark & \cmark & \cmark &    & (\cmark) \\ \midrule
w-MOEA/D \cite{szlapczynski2023ship} & Evolutionary             & \cmark & \cmark & \cmark & \cmark & \cmark &    & \\
WRM \cite{Grandcolas2022}         & Evolutionary               & \cmark & \cmark & \cmark & \cmark &        & \cmark & \\
HNDS-MPSO \cite{Zhao2022}         & Evolutionary               &         & \cmark & \cmark & \cmark & \cmark &   & \\ \midrule
Hybrid Search \cite{Precioso2024} & Evol. + Var. & \cmark  & \cmark  &         &  &  \cmark &    & (\cmark) \\
\name{} (ours)                & Evol. + Var. & \cmark & \cmark & \cmark & \cmark & \cmark & \cmark  & (\cmark) \\
\bottomrule
\end{tabular}
\caption{Representative weather-routing algorithms grouped by methodological family. A checkmark means that the method includes a mechanism directed at the criterion; it is not a formal guarantee. \cmark$^\star$: solutions are optimal on the graph, but not necessarily for the continuous problem. (\cmark): local optimality is pursued through FMS refinement (see Sections~\ref{sec:fms-second-order} and~\ref{sec:results-real}).}
\label{tab:state-of-the-art}
\end{table*}

The first family of methods, \textbf{variational algorithms}, is particularly effective in modelling the continuous nature of weather and maritime geography. For instance, the discrete Newton-Jacobi (DNJ) algorithm introduced by Ferraro \textit{et al.} \cite{Ferraro2021} demonstrated strong performance in synthetic vector fields, minimizing travel time under constant speed or reducing fuel consumption while meeting prescribed arrival times. This novel method was later renamed the Ferraro-Martín de Diego-Sato (FMS) algorithm and successfully applied to real-world optimization scenarios \cite{Jimenez2024}. Variational methods are well suited to refining a route locally when the inputs are smooth: they drive it toward a stationary point of the discretized cost, which is typically, but not necessarily, a local minimum. However, they are less effective in scenarios that require the imposition of additional constraints, such as obstacle avoidance. Moreover, incorporating complex cost functions can be challenging, particularly when derivatives are difficult or impossible to compute.

\textbf{Graph-based methods}, such as VISIR-2 \cite{mannarini2024visir}, PRM \cite{Charalambopoulos2023}, and SIMROUTE \cite{Grifoll2022}, take a different approach by discretizing the search space into a graph structure. These methods naturally excel in obstacle avoidance, as hazards can be directly accounted for during graph construction. A well-designed heuristic can also ensure that these algorithms find the best solution within the graph's confines. However, the discretized nature of these methods limits their ability to capture the full complexity of continuous environments, e.g. for modelling smooth vessel turns. Hybrid approaches, such as HADAD \cite{Jimenez2024}, have addressed this limitation by incorporating variational algorithms \cite{Ferraro2021}. Even so, graph-based methods often struggle with enforcing global constraints, such as just-in-time arrivals, because of the incremental way in which graphs are constructed. \textbf{Dynamic programming} belongs naturally to this broader family when the routing problem is discretized into a finite state space and solved through Bellman-style cost-to-go recursions over positions, headings, speeds, times, or their combinations.

\textbf{Evolutionary algorithms} offer another alternative. These methods, exemplified by w-MOEA/D \cite{szlapczynski2023ship}, WRM \cite{Grandcolas2022}, and HNDS-MPSO \cite{Zhao2022}, operate by generating a population of candidate solutions and iteratively refining them to favour lower-cost routes. Their strength lies in their adaptability: they can handle dynamic conditions, flexible cost functions, and additional constraints like obstacle avoidance or engine load through penalization strategies. However, because these methods rely on stochastic search processes, they cannot guarantee that the solution found will be locally optimal. The Hybrid Search strategy proposed by \cite{Precioso2024} attempts to bridge this gap by using an evolutionary algorithm for broad exploration, followed by the FMS algorithm \cite{Ferraro2021} to refine solutions toward a stationary point of the discrete action. Despite their flexibility, evolutionary methods often face challenges in handling complex obstacle configurations or strictly adhering to arrival constraints, which may cause them to fail to converge on valid solutions \cite{Precioso2024}.

Although these algorithmic families have significantly advanced weather routing, no single method currently satisfies all the criteria described above. Table~\ref{tab:state-of-the-art} summarizes a representative set of algorithms from each family together with the properties most relevant to our discussion. The table is not intended as an exhaustive survey; rather, it is designed to highlight the methodological gap that remains in the literature. Addressing this gap is essential for developing robust weather-routing solutions that meet the operational demands of modern maritime logistics, including efficiency, safety, and environmental sustainability.

This paper introduces \name{}, a hybrid weather routing algorithm evaluated against all seven criteria in synthetic and real-world benchmarks. Building on the exploratory flexibility of evolutionary methods and their capacity to handle various cost functions and constraints, \name{} incorporates a variational refinement step, as proposed in previous work \cite{Precioso2024}. Specifically, \name{} combines the Covariance Matrix Adaptation Evolution Strategy (CMA-ES) with the FMS method. CMA-ES is used to generate control points that define B\'ezier curves, representing candidate routes for the weather routing problem. To the best of our knowledge, this study is the first to use B\'ezier-curve parametrization for just-in-time weather routing. These candidates are then refined using the FMS method, which seeks to reduce the discrete action and approach a stationary point via Newton-type updates. In the tested settings, this hybrid approach produces competitive routes while supporting objectives such as constant engine load or Just-in-Time arrival and accommodating additional constraints through penalties and update safeguards.

\name{} participated in the SWOPP3 benchmark comparison~\cite{swopp3}, where 10 teams optimized 366 daily departures across 2024 for two trans-oceanic journeys. Methodologies included dynamic programming (6 teams), genetic algorithms (2), and isochrones (1); \name{} was the only hybrid evolutionary-variational method. Rankings were based on total energy consumption. \name{} ranked within 1.1\% of the winner, and reported the lowest computation time by over an order of magnitude.

\begin{figure*}[htbp]
    \centering
    \includegraphics[width=.8\linewidth]{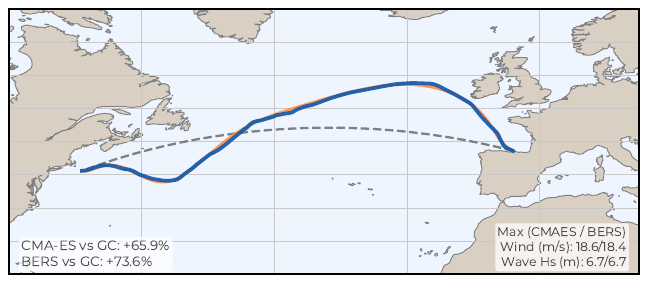}
    \caption{Representative Trans-Atlantic spring route comparing great-circle (GC), CMA-ES, and the final \name{} route (CMA-ES + FMS). The refinement from CMA-ES to \name{} is visually modest but operationally relevant, increasing the energy savings relative to the great-circle route of the same propulsion mode from 66\% to 74\% on this departure.
 Both CMA-ES and \name{} stay below the weather thresholds of wind speed (20 m/s) and wave height (7 m), although these are treated as soft constraints in the optimizer.}
    \label{fig:teaser-atlantic}
\end{figure*}

\name{} has been implemented in Python 3.12 and tested on a variety of vector fields from the literature, as well as assessed on real-ocean benchmark scenarios. All software and scripts required to regenerate the results are open-source in the public repository\footnote{\url{https://github.com/IEResearchDatalab/routetools}}, including the data-download instructions, vessel-consumption models, synthetic benchmarks, CMA-ES and FMS optimization tools, experiment configurations, validation scripts, and table and figure generators. The real-ocean experiments follow the SWOPP3 benchmark setup: Section~\ref{sec:rw-setup} describes the corridor definitions and environmental forcing, and \ref{sec:appendix-performance-model} gives the analytical model used by the open reproduction workflow. While full operational deployment of \name{} remains future work, the experiments assess all seven criteria in the tested settings. A representative result is shown in Figure~\ref{fig:teaser-atlantic}.

The remainder of this paper is organized as follows. Section~\ref{sec:algorithm} presents the details of our proposed optimization approach: B\'ezier curve parametrization, CMA-ES evolution, refinement with FMS and obstacle avoidance. Section~\ref{sec:synthetic-vectorfields} introduces the benchmark optimization problems and the results of \name{} in these vector fields. Section~\ref{sec:results-real} presents real-ocean experiments on two trans-oceanic corridors with a wind-assisted cargo vessel, demonstrating energy savings of up to 75\% from combined route optimization and wingsail technology. Finally, Section~\ref{sec:discussion} discusses the results, limitations, and practical implications. The appendices collect the canonical cost formulations, synthetic-field and land-generation details, and the reconstructed vessel model.


\section{\name{} Algorithm}
\label{sec:algorithm}

We propose a novel hybrid optimization framework that combines a derivative-free evolutionary optimizer with a continuous trajectory representation for vessel navigation. Unlike traditional stepwise pathfinding approaches, our method employs global route generation followed by iterative refinement. Specifically:

\begin{enumerate}
    \item We choose a low-dimensional parametrization of our solution space (vessel trajectories), namely by the set of 2D B\'ezier curves with $K$ control points. We apply the Covariance Matrix Adaptation Evolution Strategy (CMA-ES)~\cite{hansen2003} to explore that space and obtain a competitive candidate within that curve family.
    \item We refine the best CMA-ES candidate by re-parametrizing it as a polyline with $L$ waypoints (where $L \gg K$) and applying the Ferraro-Mart\'in de Diego-Sato (FMS) variational method~\cite{Ferraro2021}. This stage seeks further descent in the higher-dimensional waypoint space while preventing increases in the sampled land-violation count.
\end{enumerate}

The pipeline is designed around the seven criteria of Section~\ref{sec:introduction}; Table~\ref{tab:criteria} lists where each is assessed. The following subsections describe each component and the criteria it addresses.


\subsection{Parametrization with B\'ezier Curves} \label{sec:bezier}

Two-dimensional B\'ezier curves with $K$ control points have degree $K-1$ and are defined by $\pmb{c}_0, \dots, \pmb{c}_{K-1} \in \mathbb{R}^2$~\cite{bezier1972numerical}. The endpoints $\pmb{c}_0$ and $\pmb{c}_{K-1}$ correspond to the source and destination, leaving $K-2$ intermediate control points free for optimization. The general formula for a B\'ezier curve is~\cite{farin2001curves}

\begin{equation}
B(r) = \bigl(x(r), y(r)\bigr) 
= \sum_{k=0}^{K-1} \binom{K-1}{k}\,(1-r)^{K-1-k} \, r^k \,\pmb{c}_k,\quad 0 \leq r \leq 1,
\label{eq:bezier}
\end{equation}

\noindent where the parameter $r$ governs the progression along the curve, with $r=0$ corresponding to the source and $r=1$ to the destination. In weather routing applications, $r$ can be re-parametrized as travel time $t$, allowing a complete spatio-temporal description of the route; this will be particularly relevant in defining the cost function in Section~\ref{sec:cost-function}. A key advantage of this representation is that it allows one to generate a trajectory with arbitrarily fine resolution by adjusting the number of waypoints along the curve. This continuous formulation removes the dependency on fixed-step routing and enables dense intra-segment sampling, which is essential to avoid missing currents, hazards, or penalty contributions between waypoints, thus addressing Goal~\ref{cr:continuous-space}.

Despite its conceptual simplicity, the direct polynomial form of a B\'ezier curve poses numerical challenges for higher-degree curves, as the computation involves powers, binomial coefficients, and floating-point arithmetic that can accumulate rounding errors. To mitigate such instabilities, we adopt De Casteljau's algorithm~\cite{decasteljau1985mathematiques, Boehm1999}, which incrementally constructs the curve via linear interpolations of control points:

\begin{equation}
\pmb{c}_k^{(i)} := 
\begin{cases}
\pmb{c}_k^{(i-1)}\,(1-r) + \pmb{c}_{k+1}^{(i-1)}\,r, & i > 0,\\
\pmb{c}_k, & i = 0,
\end{cases}
\label{eq:casteljau}
\end{equation}

\noindent and $B(r) = \pmb{c}_0^{(K-1)}$. This recursion uses only convex combinations of points and is therefore numerically stable. Since the curve is polynomial in $r$, the resulting routes are smooth.

We implemented B\'ezier parametrization in Python using \texttt{JAX}~\cite{jax2018github} for vectorized operations and just-in-time compilation, allowing us to reshape batched control points and process multiple trajectories concurrently with minimal overhead. In this setup, users of \name{} can specify the following parameters to balance complexity, expressiveness, and computational effort.

\begin{itemize}
    \item \textbf{Number of B\'ezier control points ($K$).} A curve with $K$ control points has degree $K-1$ and $K-2$ free intermediate points. Increasing $K$ leads to more expressive trajectories, but slows the evolutionary algorithm and increases the risk of local optima.
    \item \textbf{Number of pieces ($C$).} Splitting a route into several B\'ezier pieces provides finer local control without high-degree polynomials. All experiments reported here use a single piece ($C=1$).
    \item \textbf{Number of waypoints ($L$).} We sample the curve using $L$ points evenly spaced in the interval $[0, 1]$. Higher values of $L$ define more precise trajectories at the cost of additional computational effort.
\end{itemize}


\subsection{CMA Evolutionary Strategy}  \label{sec:cmaes}

The Covariance Matrix Adaptation Evolution Strategy (CMA-ES) is a state-of-the-art derivative-free optimization algorithm designed for continuous optimization problems \cite{hansen2016cma}. CMA-ES models the search distribution as a multivariate Gaussian, iteratively updating its mean and covariance matrix to guide the search process \cite{hansen2001completely}.

In our \name{} algorithm, the initial mean solution $\pmb{\mu}_0$ is the straight line between source and destination in the synthetic experiments and the great-circle route in the real-ocean experiments. At each iteration $\tau$, the algorithm generates $P$ candidate trajectories $\{\pmb{x}_1, \dots, \pmb{x}_P\}$ by sampling from the current Gaussian distribution $\mathcal{N}(\pmb{\mu}_\tau, \mathbf{\Sigma}_\tau)$.

Each candidate is then evaluated using the cost function defined in Section~\ref{sec:cost-function}, which considers factors such as travel time or fuel consumption. Infeasible trajectories such as those crossing land are penalized using the rules described in Section~\ref{sec:land-avoidance}. Based on these evaluations, the parameters are updated to form $\mathcal{N}(\pmb{\mu}_{\tau+1}, \mathbf{\Sigma}_{\tau+1})$, a distribution that favours the best-performing points in a way that balances exploration and exploitation of the solution space.

A key advantage of CMA-ES is that its stochastic sampling allows it to escape poor local optima \cite{hansen2003}. Its adaptive covariance matrix guides the search towards promising regions while maintaining diversity among solutions. Additionally, the batch evaluation structure of CMA-ES is particularly beneficial for the \name{} algorithm, as it enables parallel computation of hundreds of candidate trajectories simultaneously, significantly accelerating the optimization process. In Just-in-Time scenarios, each segment endpoint is visited at the same time for all trajectories in the batch, which enables efficient and cache-friendly slicing along the temporal dimension.

The following parameters can be defined to control the evolution of this algorithm:

\begin{itemize}
    \item \textbf{Population size ($P$).} The number of trajectories generated and tested at each evolutionary step. Parallelization enables simultaneous simulation of many trajectories so that the computation time grows sub-linearly with $P$.
    \item \textbf{Initial standard deviation ($\sigma_0$).} This parameter controls the spread of the initial Gaussian distribution from which control points are sampled. A higher $\sigma_0$ increases exploration by generating more diverse trajectories at the start but may slow down convergence.
    \item \textbf{Tolerance.} After each iteration, CMA-ES evaluates the reduction in the cost function. If this reduction is below the specified tolerance, the evolutionary process terminates, returning the best candidate trajectory (i.e., the one with the lowest cost).
\end{itemize}


\subsection{Cost Function} \label{sec:cost-function}

The optimizer is objective-agnostic. CMA-ES requires only route-cost evaluations; FMS additionally requires first and second derivatives of the segment cost, obtained by automatic differentiation or finite differences (\ref{sec:appendix-cost}). For a discretized trajectory $\Gamma = \{\mathbf{x}_n, t_n\}_{n=1}^{L}$, we write our base optimization target as

\begin{equation}
\mathcal{C}(\Gamma) = \sum_{n=1}^{L-1} g\!\left(\mathbf{x}_n,\mathbf{x}_{n+1};\,t_n, t_{n+1}\right),
\end{equation}

\noindent where $g$ can represent travel-time, energy, or any black-box operational metric. This modularity directly supports our design goal of a flexible cost function (Goal~\ref{cr:flexible-cost}) and is essential for the real-vessel experiments in Section~\ref{sec:results-real}, where $g$ is a non-convex, non-smooth vessel performance model that was originally available only as a compiled evaluator. Canonical formulations (time minimization under constant engine load, fixed-time fuel objective, and feasibility conditions) are provided in \ref{sec:appendix-cost}.


\subsection{Land Avoidance}
\label{sec:land-avoidance}

Land is handled differently in the two optimization stages. During CMA-ES, a route that crosses land is penalized rather than discarded:
\begin{equation}
\label{eq:land-cost}
\mathcal{J}_{\mathrm{land}}(\Gamma)=\mathcal{C}(\Gamma)+\lambda_{\mathrm{land}}\,\mathcal{P}_{\mathrm{land}}(\Gamma),
\end{equation}
where $\mathcal{C}$ is the primary objective of Section~\ref{sec:cost-function} and $\mathcal{P}_{\mathrm{land}}$ counts the segments of a densely resampled curve that touch land (\ref{sec:appendix-land-penalty}). Sampling inside each segment detects crossings that a waypoint-only check would miss. It also prevents the optimizer from escaping the penalty by stretching a segment across an obstacle (\ref{cr:continuous-space}). Because CMA-ES is derivative-free, $\mathcal{P}_{\mathrm{land}}$ need not be differentiable. We use $\lambda_{\mathrm{land}}=100$ in the synthetic experiments; the real-ocean runs use $\lambda_{\mathrm{land}}=1{,}000$ together with a coastline-proximity term (Section~\ref{sec:rw-setup}).

A graded penalty is preferable to outright rejection early in the search. In narrow passages, rejecting every infeasible candidate can remove much of the population, whereas the penalty still ranks candidates by how much land they touch. Restricting the control points to water would not be sufficient either, because a B\'ezier curve can cut across a non-convex coastline even when all its control points lie at sea. Exact feasibility therefore requires a segment-level geometric check, which \name{} applies to the final real-ocean routes (see below).

During FMS, any update that increases the sampled land-violation count is undone (Section~\ref{sec:algorithm-fms}). A route that is land-free under the sampled mask therefore stays land-free, and a route that still touches land can keep improving as long as its count does not increase. The effect of $\lambda_{\mathrm{land}}$ on the search is examined in Section~\ref{sec:results-land}.

The penalty and the rollback rule steer the search away from land. In the real-ocean experiments, feasibility with respect to the coastline is verified independently after optimization: every segment of the exported route is intersected with the Natural Earth 1:10m land polygons, with longitudes unwrapped and the polygons tiled across the antimeridian. Section~\ref{sec:rw-routes} reports this check for all 1{,}464 real-ocean routes.

\subsubsection{Weather Penalties}

In the real-ocean experiments, wind speeds above $U_{\max}=20$~m/s and significant wave heights above $H_{\max}=7$~m are penalized, not rejected. Both limits are those of the SWOPP3 benchmark. We treat them as benchmark operability limits. For context, a wind speed of 20~m/s is approximately Beaufort force~8, the range addressed by IMO guidance on adverse weather \cite{IMO_MSC1228}, and 7~m lies within the range of class-society seakeeping guidance \cite{DNVGL_seakeeping}. With $U_n$ and $H_n$ the wind speed and significant wave height on segment $n$, the weather penalty is
\begin{equation}
\label{eq:penalization-env}
\mathcal{P}_{\mathrm{env}}(\Gamma) =
\lambda_H\max_n [H_n-H_{\max}]_+^2
+\lambda_U\max_n [U_n-U_{\max}]_+^2,
\end{equation}
where $[z]_+ = \max(z,0)$ and $\lambda_H,\lambda_U>0$ are separate wave and wind weights. Taking the maximum rather than a sum makes the penalty independent of the number of route samples. The complete CMA-ES objective is
\begin{equation}
\label{eq:penalization-total}
\mathcal{J}(\Gamma)=\mathcal{C}(\Gamma)+\lambda_{\mathrm{land}}\,\mathcal{P}_{\mathrm{land}}(\Gamma)+\lambda_{\mathrm{prox}}\,\mathcal{D}(\Gamma)+\mathcal{P}_{\mathrm{env}}(\Gamma),
\end{equation}
Here $\mathcal{D}(\Gamma)=\sum_{n}(\delta_n+\varepsilon)^{-1}$ is the coastline-proximity term, where $\delta_n$ is the distance from sample $n$ to the nearest land cell, read from a precomputed Euclidean distance transform of the land mask, and $\varepsilon=1$ cell avoids division by zero. It grows as a route approaches the coast and is negligible offshore. Both $\mathcal{D}$ and $\mathcal{P}_{\mathrm{env}}$ are used only in the real-ocean runs.

FMS uses a closely related but not identical objective. Land enters FMS only through the rollback rule above, and the route-wise maximum in Equation~\eqref{eq:penalization-env} is not separable over segments. FMS, which updates one waypoint at a time, therefore uses each segment's own squared excess, which over the route acts as a duration-weighted sum. The stage-specific weights are listed in Section~\ref{sec:rw-setup}. Improvements of \name{} over CMA-ES are reported in propulsive energy (Section~\ref{sec:rw-paired}).


\subsection{Trajectory Refinement with FMS}
\label{sec:algorithm-fms}

CMA-ES returns a good route within the low-dimensional B\'ezier family. FMS~\cite{Ferraro2021} then refines it in the much larger space of individual waypoints. The CMA-ES curve is discretized into a polyline $\Gamma^{(0)}=\{\mathbf{x}_1,\dots,\mathbf{x}_L\}$ with $\mathbf{x}_n=(x_n,y_n)\in\mathbb{R}^2$. The endpoints $\mathbf{x}_1$ and $\mathbf{x}_L$ are the departure and arrival points, and the interior waypoints $\mathbf{x}_2,\dots,\mathbf{x}_{L-1}$ are free.

For each interior waypoint, FMS takes a Newton step on the discrete Euler--Lagrange equation of its two adjacent segments. With $g$ the segment cost of Section~\ref{sec:cost-function}, the proposed position $\widehat{\mathbf{x}}_n$ solves
\begin{align}
\label{eq:fms-newton}
\mathbf 0 \;=\;& D_{2}g(\mathbf{x}_{n-1},\mathbf{x}_{n}) + D_{1}g(\mathbf{x}_{n},\mathbf{x}_{n+1}) \nonumber \\
&+\bigl[D_{22}g(\mathbf{x}_{n-1},\mathbf{x}_{n}) + D_{11}g(\mathbf{x}_{n},\mathbf{x}_{n+1})\bigr]\,
\bigl(\widehat{\mathbf{x}}_{n}-\mathbf{x}_{n}\bigr),
\end{align}
where $D_i$ denotes differentiation with respect to the $i$-th argument and $D_{ij}=D_i\circ D_j$. Each waypoint uses only its own $2\times2$ Hessian block, so the step ignores the coupling between neighbouring waypoints, and all interior waypoints are updated in one sweep. The derivatives are obtained by automatic differentiation when $g$ is written in \texttt{JAX}~\cite{jax2018github}, as for the performance model of \ref{sec:appendix-performance-model}, and otherwise by finite differences with a small step $h$, for example $D_1 g(\mathbf{x}_{n-1},\mathbf{x}_n)\,\mathbf{e}_1 \approx \bigl[g(\mathbf{x}_{n-1}+h\,\mathbf{e}_1,\mathbf{x}_n)-g(\mathbf{x}_{n-1},\mathbf{x}_n)\bigr]/h$. All runs reported here use automatic differentiation, since both the synthetic objectives and the performance model are written in \texttt{JAX}.

To limit overshoot, each step is damped and then clipped coordinate-wise to the mean initial waypoint spacing:
\begin{equation}
\label{eq:fms-move}
\mathbf{x}_{n} \gets \mathbf{x}_{n} + \rho\bigl(\widehat{\mathbf{x}}_{n}-\mathbf{x}_{n}\bigr), \qquad \rho = 1-d,
\end{equation}
where $d$ is the damping parameter of the implementation. We use $d=0.5$ ($\rho=0.5$) for the synthetic benchmarks and $d=0.95$ ($\rho=0.05$) for the non-smooth real-ocean objective. These values are fixed within each experiment group.

After each sweep, the sampled land-violation count is compared with that of the previous iterate. Waypoint changes that create new violations are rolled back; if this local rollback is not enough, the previous route is kept. In the time-minimization problems, segment times are then recomputed by forward accumulation; in the fixed-time problems the schedule is fixed.

FMS keeps the best route found under its objective. A stagnation counter is reset whenever a sweep improves this best value. Refinement stops after 50 consecutive non-improving sweeps in the synthetic study and 200 in the real-ocean study, or at a maximum of 500{,}000 and 10{,}000 sweeps, respectively. The stopping rule is based on stagnation of the objective rather than on the first-order residual, and the exported route is the best iterate found. Section~\ref{sec:fms-second-order} discusses what this implies for local optimality. Because the objective includes the weather penalties, FMS can trade propulsive energy for reduced weather exposure (Sections~\ref{sec:rw-paired} and~\ref{sec:rw-violations}).

In summary, CMA-ES explores the curve family globally and FMS seeks a nearby lower-cost extremal of the discrete action (\ref{cr:local-opt}). Both stages evaluate costs and constraints on densely sampled routes, so mid-segment effects are resolved at that sampling resolution (\ref{cr:continuous-space}).
\subsection{Interpreting the Local-Optimality Criterion}
\label{sec:fms-second-order}

Criterion~\ref{cr:local-opt} retains the original goal of finding the best solution within a neighbourhood. \name{} pursues it through FMS, which seeks a lower-cost extremal of the discrete action. A stationary point may be a minimum, maximum, or saddle, and the stagnation rule of Section~\ref{sec:algorithm-fms} does not certify stationarity. For the discrete problem, a stationary route is a strict local minimum if the complete Hessian with respect to all free waypoints is positive definite.

For the quadratic fixed-time fuel objective, the continuous velocity-Hessian is the identity. For the time-minimization objective, the Lagrangian is positively 1-homogeneous in velocity, so its velocity-Hessian is singular along the direction of motion; the positive-definite object is the associated Zermelo--Randers metric, which is a genuine Finsler metric when the drift is strictly slower than the vessel, $\lVert\mathbf{w}\rVert<S$~\cite{BaoRoblesShen2004}. These are pointwise Legendre-type conditions: they do not exclude conjugate points and are not a substitute for testing the complete Hessian of the discrete action with respect to all free waypoints.

We therefore assess Criterion~\ref{cr:local-opt} through the objective reduction produced by FMS in the synthetic ablation (Section~\ref{sec:ablation}) and the paired real-ocean experiments (Section~\ref{sec:rw-local-opt}). These results demonstrate systematic local improvement; they are not a general guarantee of local optimality.


\section{Validation on Synthetic Vector Fields} \label{sec:synthetic-vectorfields}

We evaluated \name{} on five well-established vector fields from the literature, each exhibiting distinct characteristics that will be detailed in this section. The goal for this section is to stress-test the seven criteria stated in the introduction. The experiments are grouped by the mechanism they exercise, and each maps onto specific criteria as recorded in Table~\ref{tab:criteria}: the open-water benchmarks of Section~\ref{sec:results-literature} address spatial continuity, dynamic conditions, cost-function flexibility and just-in-time arrival (\ref{cr:continuous-space}, \ref{cr:dynamic-conditions}, \ref{cr:flexible-cost}, \ref{cr:jit}); the ablation of Section~\ref{sec:ablation} isolates the contribution of each stage and the constant-engine-load setting (\ref{cr:constant-power}, \ref{cr:local-opt}); and the obstacle experiments of Section~\ref{sec:results-land} address land avoidance and the penalty design that supports it (\ref{cr:obstacle-avoidance}). We reserve real-world feasibility for Section~\ref{sec:results-real}. Table~\ref{tab:vectorfields} provides a summary of the synthetic vector fields, the corresponding optimization objectives, and the results obtained.

\begin{table*}[htbp]
\centering
\caption{Synthetic current environments employed to benchmark the performance of the \name{} framework across various routing scenarios and objectives. The table contrasts the best-known results from prior studies with those achieved by \name{}, as detailed in Section~\ref{sec:synthetic-vectorfields}. Computation times are for the best-performing $(\sigma_0, K)$ configuration of each field on the hardware of Section~\ref{sec:results-literature}; runtimes at the single common configuration used for the ablation and dispersion studies are reported in Sections~\ref{sec:results-dispersion} and~\ref{sec:ablation}.}
\label{tab:vectorfields}
\scriptsize
\setlength{\tabcolsep}{3pt}
\resizebox{\textwidth}{!}{%
\begin{tabular}{lllrrlrrr}
    \toprule
    \textbf{Vector} & & \textbf{Time} &  & & \textbf{Cost} & \textbf{Ref.} & \textbf{\name{}} & \textbf{Comp.} \\
    \textbf{Field} & \textbf{Formula} & \textbf{dependent?} & \textbf{Origin} & \textbf{Goal} & \textbf{Func.} & \textbf{Cost} & \textbf{Cost} & \textbf{Time (s)} \\
    \midrule
    Circular \cite{Techy2011} & Eq.~\eqref{eq:circular} & No & $\left( \cos\left( \frac{\pi}{6} \right), \sin\left( \frac{\pi}{6} \right) \right)$ & $(0, 1)$ & Time & $1.98$ & $1.98$ & 3 \\
    Four Vortices \cite{Ferraro2021} & Eq.~\eqref{eq:four-vortices} & No & $(0, 0)$ & $(6, 2)$ &  Time & $8.95$ & $8.95$ & 23 \\ 
    Double Gyre \cite{Shadden2005} & Eq.~\eqref{eq:double-gyre} & Yes & $(1.5, 0.5)$ & $(0.5, 0.5)$ & Time & $1.01$ & $0.99$ & 25 \\
    Techy \cite{Techy2011} & Eq.~\eqref{eq:techy} & Yes & $\left( \cos\left( \frac{\pi}{6} \right), \sin\left( \frac{\pi}{6} \right) \right)$ & $(0, 1)$ & Time & $1.03$  & $1.03$ & 16 \\
    Swirlys \cite{Ferraro2021} & Eq.~\eqref{eq:swirlys} & No & $(0, 0)$ & $(6, 5)$ & Fuel & $5.73$ & $1.97$ & 214 \\
    \bottomrule
\end{tabular}
}
\end{table*}

Although synthetic, these benchmarks are conservatively constructed to be challenging. Vector fields such as Four Vortices include strong counterflows that limit manoeuvrability, particularly when the surrounding currents approach the magnitude of the vessel's speed. In parallel, Perlin-generated landmasses introduce complex topological structures, often more intricate than those encountered along real-world coastlines.


\subsection{Definition of Synthetic Vector Fields}

The mathematical definitions of the five synthetic vector fields used for validation are detailed in \ref{sec:appendix-synthetic-fields}. These fields, adopted from prior literature, are designed to test algorithm performance under various conditions, including rotational flows, multiple vortices, and time-dependent dynamics.

\subsection{Land Generation} \label{sec:land}

To evaluate obstacle handling under diverse topologies, we generate synthetic coastlines with controlled complexity using Perlin noise. In the main text we only retain the benchmark complexity levels (Table~\ref{tab:land-complexity}), which are sufficient to interpret the routing results. The full procedural generation pipeline, interpolation details, and reproducibility settings are provided in \ref{sec:appendix-land-generation}.

\subsection{Results on Synthetic Vector Fields} \label{sec:results-literature}

For our implementation, we used the \texttt{pycma} library in Python \cite{hansen2019pycma}, and its function \texttt{CMAEvolutionStrategy}. All experiments were run using Python 3.12 and \texttt{JAX} 0.4.33 on an NVIDIA RTX A6000 GPU with CUDA 12.6. For reproducibility, all library versions are publicly documented in our shared repository.

We start by testing \name{} on a set of synthetic vector fields previously explored in the literature (Table~\ref{tab:vectorfields}). These cases provide a baseline to check whether \name{}, in the absence of obstacles, can match or improve upon existing methods. This also helps evaluate the performance of CMA-ES and FMS independently.

For this analysis, we focus on the effect of two parameters: the number of B\'ezier control points $K$, which controls how complex the resulting curve can be, and the initial standard deviation $\sigma_0$ of CMA-ES. The rest of the parameters are fixed, since their influence on the optimization is mostly straightforward. We fix the number of waypoints at $L=200$ to ensure curves remain smooth and continuous. The CMA-ES population size is set to $P=500$, which we found large enough to work well without significantly increasing computational cost. Algorithm tolerances are set low enough to match the precision of the results reported in the literature. Table~\ref{tab:parameters} lists the full parameter setup.

\begin{DIFnomarkup}
\begin{table}[htbp]
\centering
\begin{tabular}{ll}
\toprule
\textbf{B\'ezier configuration} & \\
N. control points, $K$ & \{3, 6, 9, 12, 15, 18\} \\
N. pieces, $C$ & 1 \\
N. waypoints, $L$ & 200 \\
\midrule
\textbf{CMA-ES configuration} & \\
Population size, $P$ & 500 \\
Init. std, $\sigma_0$ & \{0.5, 1.0, 1.5, 2.0, 2.5, 3.0\} \\
Tolerance & $1\mathrm{e}{-3}$ \\
\midrule
\textbf{FMS configuration} & \\
Damping & 0.5 \\
Patience (non-improving sweeps) & 50 \\
Maximum sweeps & 500{,}000 \\
\bottomrule
\end{tabular}
\caption{Parameter configurations used in the experimental evaluation of \name{}, for every one of the five vector fields described in Section~\ref{sec:synthetic-vectorfields}.}
\label{tab:parameters}
\end{table}
\end{DIFnomarkup}

\begin{figure}[htbp]
    \centering
    \includegraphics[width=\linewidth,trim={1.8cm 0.2cm, 1.2cm, 0.9cm},clip]{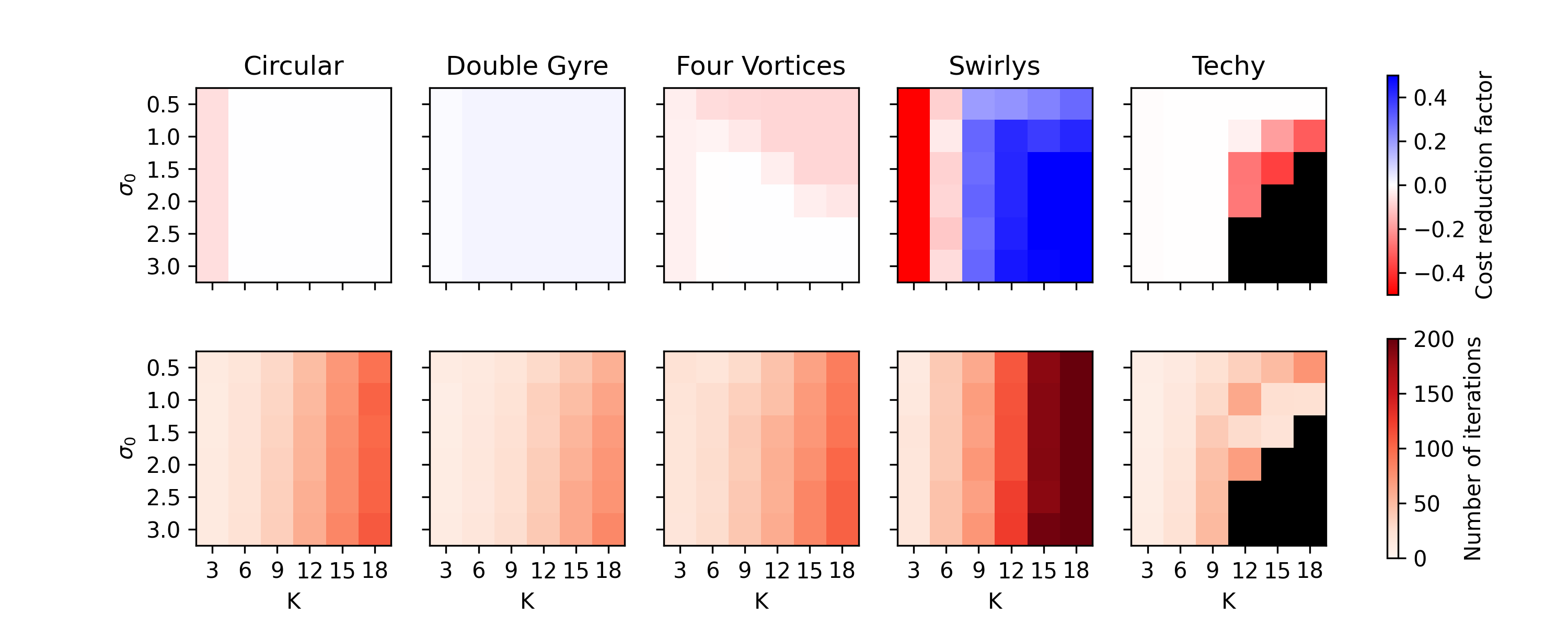}
    \caption{Cost reduction achieved by CMA-ES compared to the literature, and iteration count to convergence, as a function of initial standard deviation $\sigma_0$ and control points $K$. Black cells indicate non-convergent runs.}
    \label{fig:parameter-sensitivity-cmaes}
\end{figure}

We run the full \name{} five times on each vector field, each time using a different random seed for CMA-ES. Figure~\ref{fig:parameter-sensitivity-cmaes} shows how many iterations each configuration needed to converge (on average), and how much cost reduction it achieved compared to the literature. A reduction of zero (white in the colour maps) is already a good outcome: it means CMA-ES matches the best known result. Blue regions indicate better performance, where the method found improved solutions. Red regions are configurations that struggled, while black cells indicate cases where the optimizer couldn't even converge to a feasible solution, typically due to strong currents blocking progress.

Looking at Figure~\ref{fig:parameter-sensitivity-cmaes}, a few patterns stand out. First, $\sigma_0$ has a small effect on the required number of iterations. This suggests that CMA-ES can effectively modulate $\sigma$ across generations and populations are large enough to cover the space regardless of starting spread. On the other hand, $K$ has a clear impact: more control points mean more parameters to optimize (two per point), which increases the search space. It also affects the quality of the solution, although that depends heavily on the specific vector field. Circular, Double Gyre, and Swirlys benefit from higher $K$ values, while Four Vortices and Techy actually perform worse. For Techy in particular, increasing $K$ makes convergence harder. We found that setting $\sigma_0 = 2.0$ and $K = 9$ gave the best overall performance across the board.

\begin{figure}[htbp]
\centering
\includegraphics[width=0.9\linewidth,trim={1.8cm 0.2cm, 1.2cm, 0.9cm},clip]{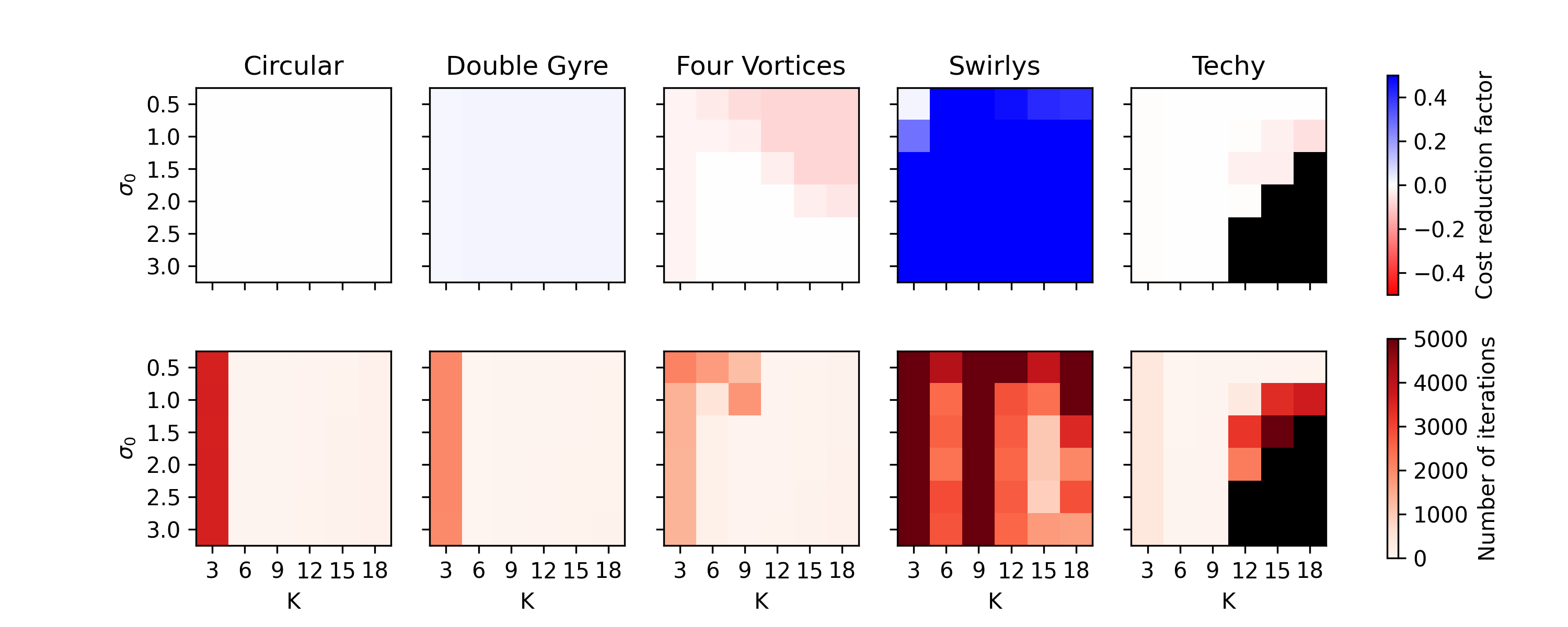}
\caption{FMS gain ($\Delta$ cost improvement) as a function of initial CMA-ES quality. Larger gains occur when CMA-ES alone performs poorly.}
\label{fig:parameter-sensitivity-fms}
\end{figure}

Figure~\ref{fig:parameter-sensitivity-fms} illustrates that FMS yields the greatest benefit when CMA-ES outputs are of lower quality, indicating that it acts as a convergence equalizer across poor initial guesses.

\subsubsection{Dispersion Across Random Seeds}
\label{sec:results-dispersion}

CMA-ES is a stochastic optimizer, so a single run is not a sufficient characterization of its behaviour. Table~\ref{tab:dispersion} reports the dispersion of the final \name{} cost across the five seeds, at the selected configuration $\sigma_0 = 2.0$, $K = 9$.

\begin{table}[htbp]
\centering
\caption{Dispersion of the final \name{} cost across five CMA-ES random seeds, at $\sigma_0 = 2.0$ and $K = 9$. Reference costs are the best published values (Table~\ref{tab:vectorfields}).}
\label{tab:dispersion}
\small
\begin{tabular}{lrrrrr}
\toprule
\textbf{Vector field} & \textbf{Ref.\ cost} & \textbf{Mean} & \textbf{Std} & \textbf{Min} & \textbf{Max} \\
\midrule
Circular      & 1.98 & 1.9750 & $<10^{-4}$ & 1.9750 & 1.9750 \\
Four Vortices & 8.95 & 8.9495 & $<10^{-4}$ & 8.9495 & 8.9495 \\
Double Gyre   & 1.01 & 0.9888 & $<10^{-4}$ & 0.9888 & 0.9888 \\
Techy         & 1.03 & 1.0317 & $<10^{-4}$ & 1.0317 & 1.0318 \\
Swirlys       & 5.73 & 1.9702 & 0.0010     & 1.9690 & 1.9711 \\
\bottomrule
\end{tabular}
\end{table}

On four of the five fields the observed seed-to-seed standard deviation is below $10^{-4}$, so the five runs agree to four decimal places; the spread is smaller than the precision at which the reference costs are published. On Swirlys, the only field with a fixed-time fuel objective and the only one where \name{} departs substantially from the published value, the standard deviation is $0.001$, or $0.05\%$ of the mean. These five seeds show small dispersion in the tested configurations, but they do not exclude rare failures or provide a high-confidence tail estimate. The pattern should not be extrapolated to the real-ocean setting of Section~\ref{sec:results-real}, where the cost landscape is non-smooth and the initialization is deliberately local ($\sigma_0=0.3$ around the great circle; Section~\ref{sec:rw-setup}).


\subsection{Ablation Study: Component Effectiveness}
\label{sec:ablation}

We compare the performance of CMA-ES alone (using $\sigma_0 = 2.0$ and $K = 9$), FMS alone, and their combined use in \name{}. This serves as an ablation study to highlight the necessity of the two-stage architecture; the FMS-from-straight-line case is not intended as a competitive baseline but rather to illustrate the behaviour of the variational method with a naive initialization. In the original work that introduced FMS \cite{Ferraro2021}, multiple random initial guesses were used. Here, we opt for the straight-line path as a simple and reproducible baseline, and 200 waypoints to match the CMA-ES resolution.

\begin{table}[htbp]
\centering
\begin{tabular}{lrrrr}
    \toprule
    \textbf{Vector} & \textbf{Ref.} & \textbf{CMA-ES} & \textbf{FMS} & \textbf{\name{}} \\
    \textbf{Field} & \textbf{Cost} & \textbf{Cost} & \textbf{Cost} & \textbf{Cost} \\
    \midrule
    Circular \cite{Techy2011} & $1.98$ & $1.98$ & $1.98$ & $1.98$ \\
    Four Vortices \cite{Ferraro2021} & $8.95$ & $8.95$ & $9.69$ & $8.95$ \\ 
    Double Gyre \cite{Shadden2005} & $1.01$ & $0.99$ & $0.99$ & $0.99$ \\
    Techy \cite{Techy2011} & $1.03$ & $1.03$ & $1.04$ & $1.03$ \\
    Swirlys \cite{Ferraro2021} & $5.73$ & $3.39$ & $5.59$ & $1.97$ \\
    \bottomrule
\end{tabular}
\caption{Results on synthetic vector fields comparing CMA-ES, FMS, and the full \name{} pipeline.}
\label{tab:results-literature}
\end{table}

In general, CMA-ES and FMS both perform close to the reference values, particularly on simpler vector fields. CMA-ES alone takes approximately 1 second to converge across all test cases, typically requiring 50 generations with a population of $P=500$ individuals per generation, yielding $P \times G = 25{,}000$ total function evaluations (parallelized within each generation). In contrast, FMS initialized from a straight line can take up to 40 seconds and roughly 50{,}000 function evaluations. These evaluations are sequential and therefore more computationally demanding.

When combining both stages in \name{}, the output of CMA-ES already provides a good-quality path, which significantly reduces the refinement effort. On the four time-minimization fields, FMS then converges in fewer than 100 iterations and the total runtime of the full pipeline is under 5 seconds (mean $1.4$--$3.2$~s across seeds). Swirlys is the exception: its fixed-time fuel objective requires substantially more FMS sweeps, and the runtime is both larger and far more variable across seeds ($39 \pm 40$~s), which is the only appreciable source of run-to-run variability we observe in the synthetic benchmarks. The most noticeable improvements in solution quality are likewise found in the more complex vector fields, particularly Swirlys, where \name{} clearly outperforms both algorithms performing on their own.

The ability of \name{} to match or exceed performance on Circular and other steady-state vector fields validates its handling of fixed propulsion constraints, a key aspect of Goal~\ref{cr:constant-power}.

\begin{figure}[htbp]
    \centering
    \begin{subfigure}{0.4\textwidth}
        \includegraphics[width=\linewidth,trim={0 0.5cm, 0, 0.9cm},clip]{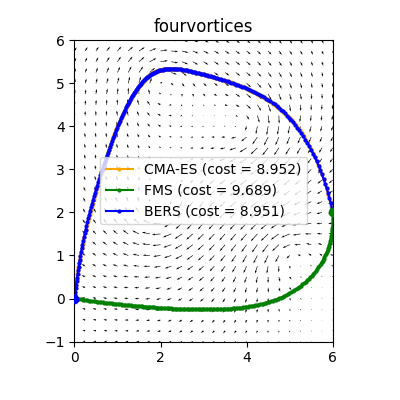}
        \caption{Four Vortices synthetic field \eqref{eq:four-vortices}.}
        \label{fig:results-literature-fourvortices}
    \end{subfigure}
    \begin{subfigure}{0.4\textwidth}
        \includegraphics[width=\linewidth,trim={0 0.8cm, 0, 0.92cm},clip]{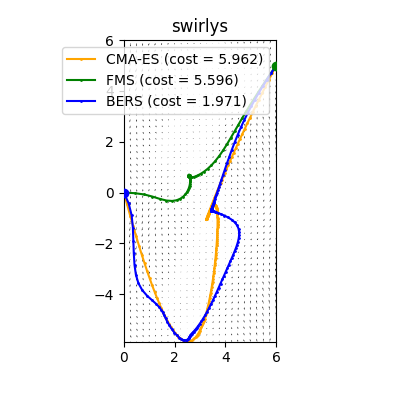}
        \caption{Swirlys vector field under a fuel minimization objective with fixed arrival time $T = 30$.}
        \label{fig:results-literature-swirlys}
    \end{subfigure}
    \caption{Routes obtained by the CMA-ES algorithm alone (orange), by the FMS algorithm initialized with a straight line (green), and by \name{} (blue). The background arrow map shows the vector field speed and direction.}
\end{figure}

Figure~\ref{fig:results-literature-fourvortices} shows the difference in behaviour between CMA-ES and FMS on the Four Vortices field. CMA-ES explores the search space more broadly, while FMS, starting from a straight line, tends to converge to a nearby local solution. When combined, the full \name{} pipeline takes advantage of both strategies and recovers the best result reported in the literature. This is consistent with \ref{cr:continuous-space}, since route costs are evaluated on dense discretizations that resolve strong mid-segment gradients.

The Double Gyre and Techy vector fields introduce time dependence, with evolving flow structures over the course of the route. \name{} matches or improves the results reported in the literature, demonstrating its ability to adapt to temporally varying environments (Goal~\ref{cr:dynamic-conditions}). On Techy, the \name{} cost (1.0317 across all five seeds; Table~\ref{tab:dispersion}) agrees with the published value to its reported precision. On Double Gyre, \name{} reaches 0.989, about 2\% below the published 1.01; since we have not determined whether this reflects a genuinely better route or differences in discretization and time integration between implementations, we do not claim it as a performance gain.

The Swirlys vector field is particularly challenging due to its fixed travel-time constraint ($T = 30$). This scenario is a strong test of Goal~\ref{cr:jit}, as the optimizer must adhere to the schedule while minimizing energy use. Figure~\ref{fig:results-literature-swirlys} shows how \name{} successfully navigates this trade-off by adjusting trajectory shape and timing. In this case, \name{} identifies a more energy-efficient path than that reported by Ferraro \textit{et al.}~\cite{Ferraro2022}. It is worth noting, however, that the authors' objective in that work was to demonstrate the general convergence behaviour of their algorithm, so a direct comparison of the final cost has limited meaning.

Additionally, the Swirlys case uses a fuel consumption model instead of a traditional time-minimization objective. \name{}'s ability to handle this alternate cost structure, while still delivering high-quality routes, supports \ref{cr:flexible-cost}. CMA-ES accepts derivative-free objectives, while FMS can refine custom objectives for which the required branchwise derivatives can be evaluated.

With respect to Criterion~\ref{cr:local-opt}, the synthetic ablation shows the practical role of refinement without establishing a general guarantee. The full pipeline preserves the CMA-ES objective to the reported precision on the four time-minimization fields and reduces the Swirlys cost from 3.39 to 1.97 (Table~\ref{tab:results-literature}). A computer-assisted proof based on interval arithmetic further certifies a strict local minimum of the discrete problem next to each of the 20 travel-time routes (for deformations normal to the route) and, after Newton post-processing, for three of the five Swirlys runs; the certificates and code are available in the public repository.

Together, these experiments show how \name{} addresses multiple criteria of a weather-routing algorithm in the tested settings, including spatial continuity, obstacle handling, dynamic adaptation, timing constraints, and cost-function flexibility.


\subsection{Results on Land Avoidance} \label{sec:results-land}

After validating \name{} in open-water scenarios, we now test whether it can handle geographic constraints, specifically routing through environments in which landmasses block direct paths. This experiment directly targets \ref{cr:obstacle-avoidance}, which requires feasible routing in the presence of static hazards. It also probes \ref{cr:continuous-space}, because land interactions are computed through dense intra-segment sampling rather than only at waypoint locations.

To avoid oversimplification, we retain the Four Vortices vector field and superimpose land obstacles, thereby combining strong flow dynamics with spatial constraints. We define three levels of land complexity based on the land-generation parameters introduced in Section~\ref{sec:land}:

\begin{table}[htbp]
\centering
\caption{Land complexity levels used in benchmarking.}
\label{tab:land-complexity}
\begin{tabular}{lcc}
\toprule
\textbf{Complexity Level} & \textbf{Resolution} & \textbf{Water Level} \\
\midrule
Easy   & 3 & 0.9 \\
Medium & 4 & 0.8 \\
Hard   & 5 & 0.7 \\
\bottomrule
\end{tabular}
\end{table}

For each level, we generate 50 random land configurations and run the full \name{} pipeline to compute a path from $(0,0)$ to $(6,2)$. The goal is to recover a high-quality route that respects land constraints and still benefits from the vector field.

\begin{figure}[htbp]
    \centering
    \includegraphics[width=0.7\linewidth,trim={0.5cm 0.2cm 0.2cm 0.4cm},clip]{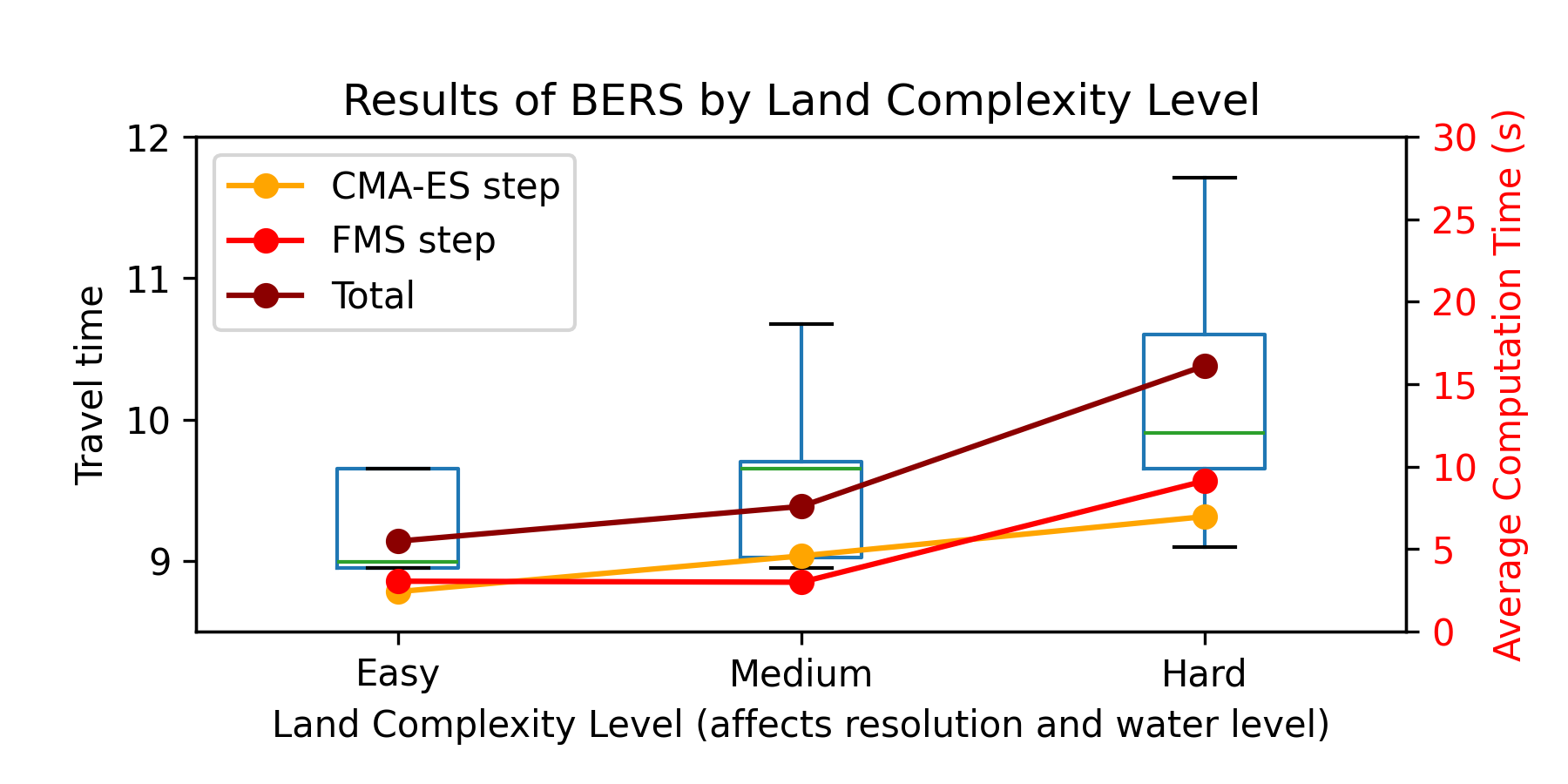}
    \caption{Performance of \name{} across increasing land complexity. Left axis shows a box-plot of travel times over 50 random instances; right axis shows average computation time, decomposed into CMA-ES and FMS steps.}
    \label{fig:results-land-complexity}
\end{figure}

Figure~\ref{fig:results-land-complexity} shows the average travel time and computation time across the three complexity settings. As expected, more land (i.e., lower water level and higher resolution) leads to longer travel times and higher computational costs. Most of the time is spent during the CMA-ES stage, which faces an increasingly difficult search problem. FMS remains fast, but its contribution grows more important as land constraints become tighter.

\begin{figure}[htbp]
    \centering
    \begin{subfigure}{0.3\textwidth}
        \includegraphics[width=\linewidth,trim={0 0.6cm 0 1.18cm},clip]{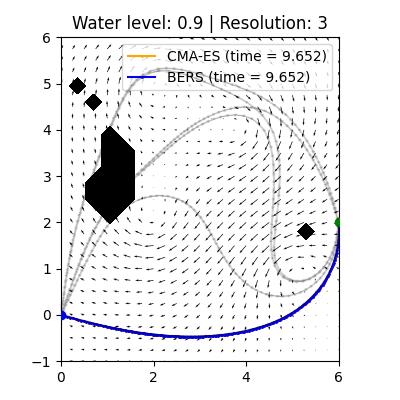}
        \caption{\textbf{Easy} scenario.}
        \label{fig:results-land-easy}
    \end{subfigure}
    \begin{subfigure}{0.3\textwidth}
        \includegraphics[width=\linewidth,trim={0 0.6cm 0 1.18cm},clip]{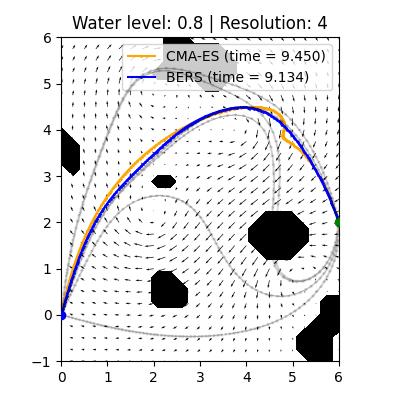}
        \caption{\textbf{Medium} scenario.}
        \label{fig:results-land-medium}
    \end{subfigure}
    \begin{subfigure}{0.3\textwidth}
        \includegraphics[width=\linewidth,trim={0 0.6cm 0 1.18cm},clip]{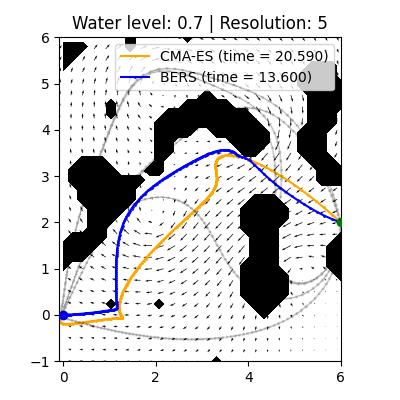}
        \caption{\textbf{Hard} scenario.}
        \label{fig:results-land-hard}
    \end{subfigure}
    \caption{Example routes obtained by \name{} (blue) and CMA-ES alone (orange) across different land complexity levels. In dim grey, the original solutions found for the no-land scenario. The background shows the vector field and land obstacles.}
    \label{fig:results-land-complexity-examples}
\end{figure}

To better understand how the two stages behave in practice, Figures~\ref{fig:results-land-easy} to~\ref{fig:results-land-hard} show three example runs, one from each complexity level. The blue curve is the final output of \name{}, while the orange curve shows the raw CMA-ES result before refinement. The background flow (Four Vortices) and land patches (black) are also shown.

In the easy setting (Figure~\ref{fig:results-land-easy}), CMA-ES already finds a clean solution, and FMS makes only minor adjustments. The grey curves indicate the locally optimal paths reported by Ferraro \textit{et al.}~\cite{Ferraro2022}. In this case, \name{} follows the only viable path previously identified that remains unobstructed. In the medium case (Figure~\ref{fig:results-land-medium}), \name{} again recovers one of the paths reported by Ferraro \textit{et al.}~\cite{Ferraro2022}. The hard case (Figure~\ref{fig:results-land-hard}) is more demanding: all previously known paths are blocked by land and \name{} must identify a new one. CMA-ES initially converges to a long detour, which FMS then corrects, recovering a trajectory much closer to the land-free optimum. The refinement step also smooths the route transitions near the land boundary.

This pattern highlights the complementary roles of both stages. CMA-ES provides global exploration when corridors are narrow or fragmented, while FMS improves curvature and removes detours introduced by optimization noise or local minima. Together, they produce results close to those reported in the literature under the sampled constraints. These examples support \ref{cr:continuous-space} by accounting for mid-segment effects through dense sampling, alongside \ref{cr:obstacle-avoidance} and \ref{cr:dynamic-conditions}, while illustrating the value of variational refinement in complex environments (\ref{cr:local-opt}).

\subsubsection{Sensitivity to the Land Penalty Coefficient}
\label{sec:results-lambda-land}

Because the land constraint is enforced through a finite penalty during the CMA-ES stage (Section~\ref{sec:land-avoidance}), the value of $\lambda_{\mathrm{land}}$ is a design parameter whose influence must be established rather than assumed. Too small a coefficient fails to exclude infeasible candidates; too large a coefficient could in principle distort the cost landscape and displace the optimum. We therefore repeated the land experiment on one representative configuration at each complexity level, sweeping $\lambda_{\mathrm{land}} \in \{10, 50, 100, 500, 1000\}$ with five random seeds per cell, and recorded the fraction of final solutions with zero violations under the same densely sampled synthetic-land mask used by the optimizer. Table~\ref{tab:lambda-land} reports the result.

\begin{table}[htbp]
\centering
\caption{Sensitivity to the land penalty coefficient $\lambda_{\mathrm{land}}$. Each cell aggregates five CMA-ES runs with different random seeds on one land configuration per complexity level. Feasibility is the fraction of final solutions with zero violations under the optimizer's densely sampled synthetic-land mask. Costs are travel times in field units; no sampled-feasible solution exists to average at $\lambda_{\mathrm{land}} = 10$ for Medium and Hard.}
\label{tab:lambda-land}
\begin{tabular}{lrrrr}
\toprule
\textbf{Scenario} & \textbf{$\lambda_{\mathrm{land}}$} & \textbf{Feasibility (\%)} & \textbf{Mean cost} & \textbf{Generations} \\
\midrule
Easy   &   10 & 100.0 & 11.05 &   9.6 \\
Easy   &   50 & 100.0 &  9.65 &  34.6 \\
Easy   &  100 & 100.0 &  9.65 &  34.0 \\
Easy   &  500 & 100.0 &  9.65 &  34.0 \\
Easy   & 1000 & 100.0 &  9.65 &  33.8 \\
\midrule
Medium &   10 &   0.0 &  ---  &  10.0 \\
Medium &   50 & 100.0 &  9.65 &  38.2 \\
Medium &  100 & 100.0 &  9.65 &  38.4 \\
Medium &  500 & 100.0 &  9.65 &  38.8 \\
Medium & 1000 & 100.0 &  9.65 &  38.6 \\
\midrule
Hard   &   10 &   0.0 &  ---  &  10.0 \\
Hard   &   50 & 100.0 &  9.76 & 321.0 \\
Hard   &  100 & 100.0 &  9.76 & 320.4 \\
Hard   &  500 & 100.0 &  9.76 & 319.8 \\
Hard   & 1000 & 100.0 &  9.76 & 320.2 \\
\bottomrule
\end{tabular}
\end{table}

The behaviour is threshold-like in these tested configurations. At $\lambda_{\mathrm{land}}=10$ no run is sampled-feasible at Medium or Hard complexity. From $\lambda_{\mathrm{land}}=50$ onwards, every run at every tested level has zero sampled violations, and the converged cost and generation count are nearly invariant up to $\lambda_{\mathrm{land}}=1000$. Thus the synthetic value $\lambda_{\mathrm{land}}=100$ lies inside a tested plateau rather than at a uniquely tuned optimum. This finding concerns one configuration per complexity level and the optimizer's mask; it is not evidence that the distinct 1{,}000 real-ocean coefficient lies on the same plateau, nor is it an exact geometric guarantee between samples. Even at Easy complexity, where $\lambda_{\mathrm{land}}=10$ is sampled-feasible, it yields a 15\% higher cost than the plateau values.

A remark on scope is warranted. This study varies $\lambda_{\mathrm{land}}$ on CMA-ES alone, so it characterizes the stage at which the penalty acts and isolates it from the FMS rollback rule. It documents a plateau in the tested configurations, not a universal guarantee: the boundary depends on the scale of the primary objective, and for an objective with a substantially different magnitude the sweep should be repeated. In practice the ratio $\lambda_{\mathrm{land}} / \mathcal{C}$ is the quantity that matters, and the values above correspond to $\lambda_{\mathrm{land}}$ between five and one hundred times the typical converged cost.


\section{Validation on Real-World Scenarios: The SWOPP3 Benchmark}  \label{sec:results-real}

Building on the synthetic validation in Section~\ref{sec:synthetic-vectorfields}, we assess \name{} on the real-ocean test bed used in SWOPP3 \cite{swopp3}. It contains two trans-oceanic routes for a cargo vessel equipped with a Wind Propulsion System~(WPS).

Route optimization and wind conditions interact in both propulsion modes. The optimizer can reduce aerodynamic drag and exploit favourable wave conditions with or without WPS. When WPS is enabled, it can also seek apparent wind angles that increase wingsail thrust.

We use the Just-in-Time formulation of Criterion~\ref{cr:jit} and evaluate every configuration over a full year of ERA5 reanalysis. The quadratic synthetic objective is replaced by the non-convex vessel-performance model used as the segment cost $g$ in Section~\ref{sec:cost-function}.

The energy values reported below are propulsive energy only. The optimizer itself minimizes a penalized objective that also contains the land and weather terms of Section~\ref{sec:land-avoidance}. FMS can therefore accept a small increase in propulsive energy when this lowers the complete penalized objective. Sections~\ref{sec:rw-paired} and~\ref{sec:rw-violations} show that trade-off.

This section provides real-ocean evidence for five criteria from Table~\ref{tab:criteria}: fixed-time arrival (\ref{cr:jit}), a non-convex cost (\ref{cr:flexible-cost}), obstacle handling against a real coastline (\ref{cr:obstacle-avoidance}), adaptation over an annual weather cycle (\ref{cr:dynamic-conditions}), and local improvement through variational refinement (\ref{cr:local-opt}). The relevant evidence appears in Sections~\ref{sec:rw-setup}, \ref{sec:rw-routes}, \ref{sec:rw-paired}, \ref{sec:rw-seasonal}, and~\ref{sec:rw-local-opt}.

\subsection{Experimental Setup}  \label{sec:rw-setup}

All real-ocean inputs used in this section are taken from the SWOPP3 benchmark setup. During the competition, the vessel performance model was distributed to participants as a compiled binary and used as a black-box evaluator. \ref{sec:appendix-performance-model} gives an open analytical reconstruction of that evaluator and quantifies its residual error against the binary, allowing independent verification of the routing--energy interface and close reproduction of the reported energy figures.

We consider a generic 88~m single-skeg cargo vessel with electric propulsion, a controllable-pitch propeller, and four 138~m$^2$ rigid wingsails. The total sail area is 552~m$^2$.

The SWOPP3 model computes net propulsive power at the midpoint of each route segment. Its inputs are vessel speed through water, true wind speed and angle, significant wave height, and mean wave angle. The model combines hull resistance, aerodynamic drag, wave-added resistance and, when WPS is enabled, wingsail thrust. The resulting power is clamped at zero. Integrating it over the segment durations gives total voyage energy in MWh.

We evaluate the vessel on the two corridors in Table~\ref{tab:corridors}: Santander to New York and Tokyo to Los~Angeles. Their fixed passage times are 354~h and 583~h, respectively, thereby enforcing Criterion~\ref{cr:jit}. Under this formulation, local route geometry and the total voyage duration determine the speed of each segment.

No explicit speed or propulsion-power caps are imposed as external constraints in the routing optimization. Instead, admissibility is bounded from two directions. First, the performance model is calibrated and validated over a fixed input domain (Table~\ref{tab:perf-domain}), of which the relevant bound is a ship speed through water of $0$--$14.5$~m/s ($0$--$28.2$~kn); evaluations outside this domain would constitute extrapolation. Second, the cubic hull-resistance term $K_H v^3$ of Equation~\eqref{eq:hull} makes high segment speeds intrinsically expensive, so the optimizer is strongly deterred from sustaining them under a fixed arrival time. Table~\ref{tab:rw-speeds} reports the resulting distribution of implied segment speeds across all 366 departures in each configuration, verifying \emph{a posteriori} that this indirect mechanism is sufficient: every one of the $343{,}308$ evaluated segments lies inside the model's validated speed domain, so no result in this section relies on extrapolation of the performance model.

Implied segment speed is computed from the exported polyline as $v_n=d(\mathbf{x}_n,\mathbf{x}_{n+1})/\Delta t_n$. Across the four configurations, the mean is $8.2$--$8.7$~kn and the central 90\% lies between $5.6$ and $11.6$~kn.

The tails are rare: 0.86\% of segments are below 3.9~kn and 0.054\% are above 15~kn. The extremes are below 0.01 and 23.1~kn. They arise because the fixed-time schedule gives every segment the same duration while neither CMA-ES nor FMS imposes a speed penalty. Uniform sampling in the B\'ezier parameter also gives the initial CMA-ES route unequal segment lengths, after which FMS moves the waypoints freely.

All values remain inside the surrogate model's validated domain. This is not an operational speed or power certification. A deployment version should impose vessel-specific limits directly or optimize an arc-length-aware time parametrization.

\begin{table}[htbp]
\centering
\caption{Trans-oceanic corridors used in the real-ocean experiments.}
\label{tab:corridors}
\begin{tabular}{lllrr}
\toprule
\textbf{Corridor} & \textbf{Origin} & \textbf{Destination} & \textbf{GC Dist.} & \textbf{Pass. Time} \\
& & & \textbf{(nm)} & \textbf{ (h)} \\
\midrule
Atlantic & Santander (ESSDR) & New York (USNYS) & 2826 & 354 \\
Pacific  & Tokyo (JPTYO) & Los~Angeles (USLAX) & 4663 & 583 \\
\bottomrule
\end{tabular}
\end{table}

\begin{table}[htbp]
\centering
\caption{Distribution of implied segment speeds (knots) of the final \name{} routes, aggregated over all 366 departures of 2024 in each configuration. All values lie within the performance model's validated speed domain of $0$--$28.2$~kn (Table~\ref{tab:perf-domain}).}
\label{tab:rw-speeds}
\small
\begin{tabular}{lrrrrrrr}
\toprule
\textbf{Configuration} & \textbf{Segments} & \textbf{Min} & \textbf{P5} & \textbf{Mean} & \textbf{Median} & \textbf{P95} & \textbf{Max} \\
\midrule
Atlantic, no WPS & 64{,}782  & 0.18 & 5.63 & 8.54 & 8.58 & 11.13 & 16.56 \\
Atlantic, WPS    & 64{,}782  & 0.37 & 5.78 & 8.70 & 8.66 & 11.56 & 19.82 \\
Pacific, no WPS  & 106{,}872 & 0.80 & 6.84 & 8.16 & 8.15 &  9.44 & 23.11 \\
Pacific, WPS     & 106{,}872 & 0.00 & 5.55 & 8.38 & 8.28 & 11.46 & 22.84 \\
\bottomrule
\end{tabular}
\end{table}

\begin{figure}[htbp]
    \centering
    \includegraphics[width=0.75\linewidth]{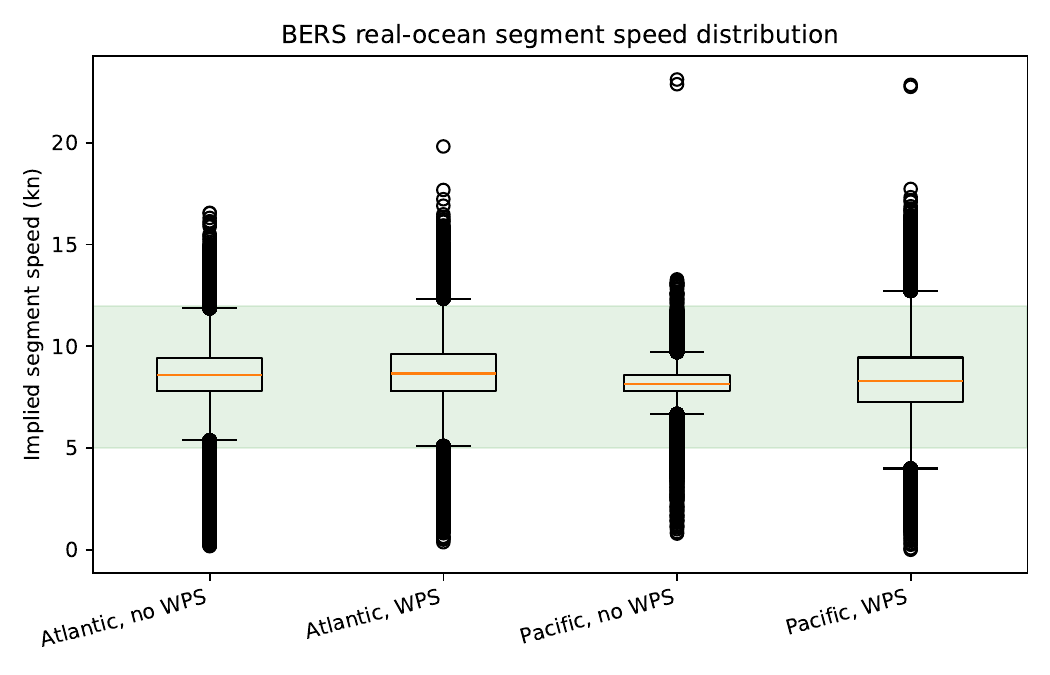}
    \caption{Distribution of implied segment speeds of the final \name{} routes for each configuration, over all 366 departures. The shaded 5--12~kn band is a visual reference spanning approximately the combined central 90\% of segments, not an imposed operational limit. Boxes span the interquartile range; whiskers extend to 1.5 times the interquartile range.}
    \label{fig:rw-speeds}
\end{figure}

Environmental forcing comes from official ERA5 reanalysis for 2024: 10~m wind, significant wave height, and mean wave direction \cite{era5}. The fields have a $0.25^\circ$ spatial grid and one-hour time steps. Values along each route are obtained by tricubic interpolation in space and time.

Wind and wave limits enter through the soft penalties of Section~\ref{sec:land-avoidance}, at 20~m/s and 7~m. For each corridor and propulsion mode, we compare the great circle, CMA-ES alone, and the full \name{} route. All four configurations include 366 daily departures at 12:00 UTC.

We report both the CMA-ES result and the final route after FMS. CMA-ES starts from the great circle. It uses 178 waypoints in the Atlantic and 293 in the Pacific, or about two hours per segment, while candidates are evaluated on a finer 30-minute grid.

The Atlantic search uses $(K,P)=(10,400)$ and the Pacific search $(15,400)$. Both use $\sigma_0=0.3$ and at most 50{,}000 evaluations. The sampled-land penalty is $10^3$ and the distance-to-land weight is 10. Weather weights are $(\lambda_U,\lambda_H)=(50,50)$ in the Atlantic and $(50,200)$ in the Pacific.

Energy use and weather exposure are competing objectives. The weighted sum above is one scalarization of that multi-objective problem. We selected the reported weights in preliminary trials near the observed elbow of the trade-off: smaller weights left more threshold violations, whereas larger weights produced substantial detours and rapidly increasing energy. Those exploratory runs were not retained as a formal sensitivity dataset, so we do not claim that these weights are uniquely optimal. A systematic Pareto-front analysis is left for future work.

FMS uses weights $(50,50)$ in both corridors. Its damping is $d=0.95$, with a patience of 200 non-improving sweeps and a maximum of 10{,}000 sweeps. The public run scripts record all settings.

As mentioned in the introduction, \name{} ranked within 1.1\% of the best-performing submission in the SWOPP3 multi-participant benchmark comparison. Our algorithm also reported the lowest computation time by more than one order of magnitude. \name{} achieved the best score among all submissions in the Pacific corridor with WPS enabled, indicating that its combination of route flexibility and local refinement is especially effective when the feasible corridor is broad and the optimization can fully exploit favourable apparent wind angles. This result is consistent with the design of \name{}, whose CMA-ES stage evaluates large batches of candidate trajectories in parallel.

The complete reproduction workflow is open-source in the public repository.\footnote{\url{https://github.com/IEResearchDatalab/routetools}} Following its instructions reproduces the results, tables, and figures reported here. The guide links each output to an executable script and explains how to download the public ERA5 \cite{era5} and Natural Earth data. The same repository contains the analytical vessel models, synthetic benchmarks, optimization tools, resolved configurations, validation code, and plotting scripts. No proprietary software or data are needed to regenerate the paper.

The open workflow uses the analytical model in \ref{sec:appendix-performance-model}. Against the historical compiled evaluator, its maximum error was 0.103~kW, or 0.031\% relative. The binary is retained only as an independent reference. Other teams' scores remain external benchmark results, so the reported 1.1\% margin and runtime comparison are not outputs of our workflow.


\subsection{Route Geometry}  \label{sec:rw-routes}

\begin{figure}[htbp]
\centering
\begin{subfigure}[t]{\linewidth}
    \centering
    \includegraphics[width=\linewidth]{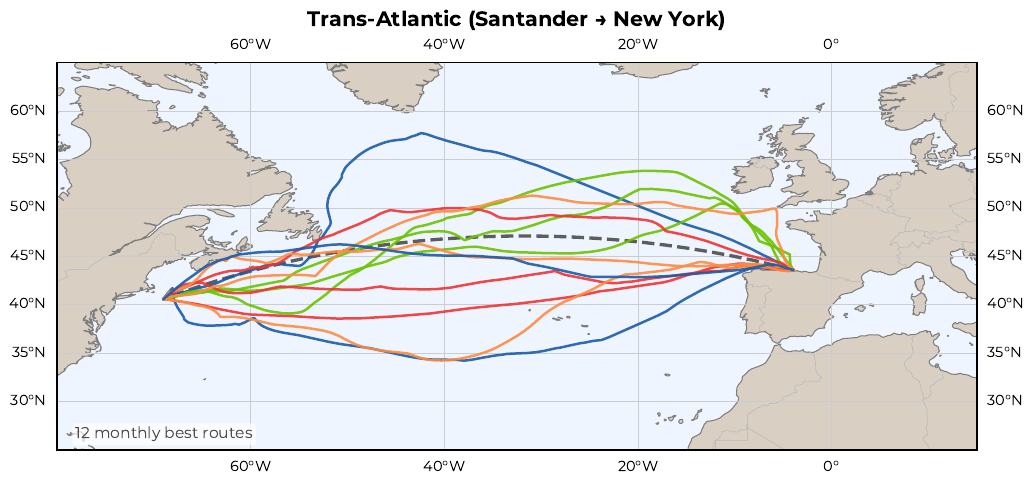}
    \caption{Trans-Atlantic corridor.}
\end{subfigure}

\begin{subfigure}[t]{\linewidth}
    \centering
    \includegraphics[width=\linewidth]{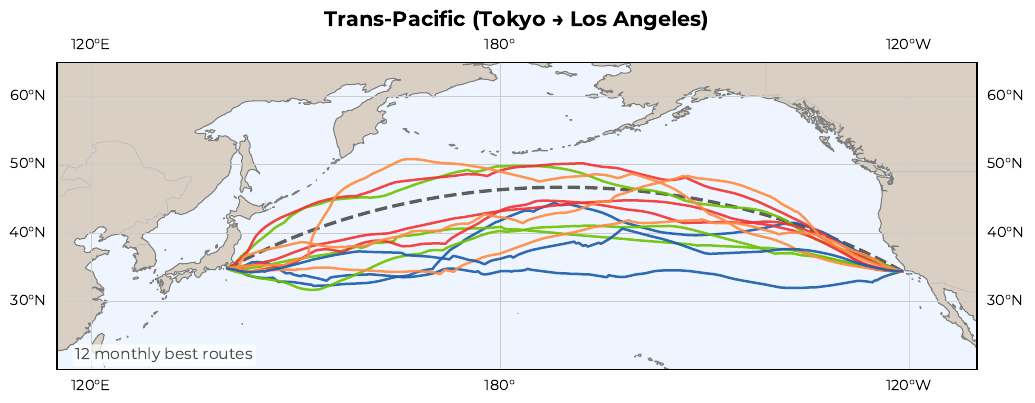}
    \caption{Trans-Pacific corridor.}
\end{subfigure}
\caption{Representative optimized routes (\name{}) versus the great-circle baseline (dashed). Departures are coloured by season: spring in green, summer in red, autumn in yellow, winter in blue.}
\label{fig:rw-routes}
\end{figure}

Figure~\ref{fig:rw-routes} shows one optimized departure per month and corridor, coloured by season. The mean \name{} route length is 3{,}078~nm in the Atlantic and 4{,}884~nm in the Pacific. The corresponding great-circle distances are 2{,}826 and 4{,}663~nm.

Winter routes deviate most from the great circle, in line with their stronger weather. Some routes also run close to the coast. An operational planner could require more clearance by adding a distance-to-coast penalty.

Because several routes pass close to land, we checked all 1{,}464 final polylines against Natural Earth 1:10m polygons. This check is independent of the raster mask used during optimization. Pacific longitudes are unwrapped and the land geometry is tiled across the antimeridian. A regression test confirms that a $179^\circ$ to $-179^\circ$ segment is treated as a short dateline crossing, not as a chord across the globe.

No segment of any of the 1{,}464 routes intersects land: the audit returns zero intersecting routes in each of the four configurations, 366 departures each.

This check complements the optimizer rather than changing the FMS claim. FMS preserves a route that is already water-only under its densely sampled mask. The polygon step confirms that agreement against the independent coastline geometry used for publication.

\subsection{Energy Savings from Route Optimization}  \label{sec:rw-energy}

\begin{figure}[htbp]
\centering
\begin{subfigure}[t]{0.4\linewidth}
    \centering
    \includegraphics[width=\linewidth,trim={0 2 0 0},clip]{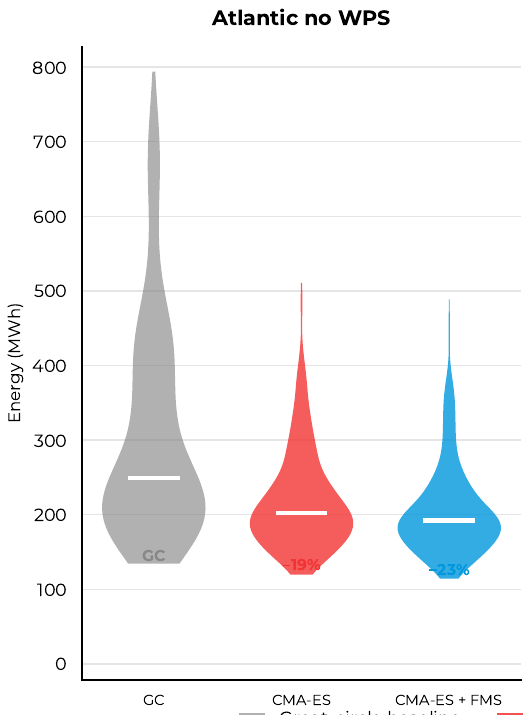}
    \caption{Trans-Atlantic, WPS disabled.}
\end{subfigure}
\begin{subfigure}[t]{0.4\linewidth}
    \centering
    \includegraphics[width=\linewidth,trim={0 2 0 0},clip]{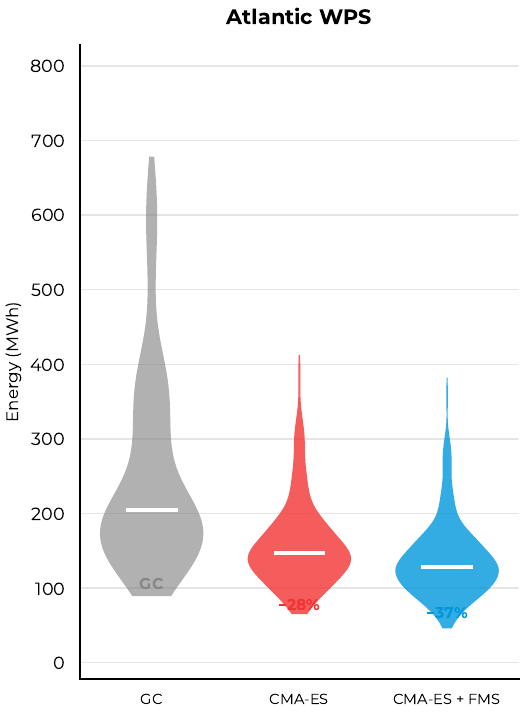}
    \caption{Trans-Atlantic, WPS enabled.}
\end{subfigure}
\caption{Voyage energy distribution (MWh) across 366 departures of 2024 for each experimental configuration. Great-circle baseline in grey, CMA-ES optimized routes in red, \name{} in blue.}
\label{fig:rw-energy-overview-atlantic}
\end{figure}

Figure~\ref{fig:rw-energy-overview-atlantic} shows the Trans-Atlantic energy distributions with and without WPS across all 366 departures. For brevity, we display the Atlantic distributions only; the Pacific corridor exhibits the same qualitative behaviour and is summarized numerically in Table~\ref{tab:rw-summary}. Two patterns are consistent across strategies: (1) great-circle (GC) routes exhibit a broader high-energy tail, indicating departures under extreme weather where adaptive routing can avoid costly exposure; and (2) enabling WPS shifts distributions downward in all strategies, with a slightly longer low-energy tail than in the no-WPS case, showing that favourable wind scenarios can produce especially large reductions.

\begin{table*}[htbp]
\centering
\caption{Mean voyage energy and energy savings for all experimental configurations across 366 departures of 2024. ``Saving vs GC\textsubscript{same}'' compares optimized to the great-circle baseline of the same WPS mode; ``Saving vs GC\textsubscript{noWPS}'' compares to the propulsion-only great-circle baseline.}
\label{tab:rw-summary}
\begin{tabular}{lllrrrr}
\toprule
\textbf{Corridor} & \textbf{WPS} & \textbf{Strategy} & \textbf{Mean} & \textbf{Std} & \textbf{Saving vs GC\textsubscript{same}} & \textbf{Saving vs GC\textsubscript{noWPS}} \\
& & & \textbf{(MWh)} & \textbf{(MWh)} & \textbf{(\%)} & \textbf{(\%)} \\
\midrule
Atlantic & Disabled & Great Circle & 310.3 & 148.7 & --- & --- \\
Atlantic & Disabled & CMA-ES       & 220.2 & 67.0 & 29.0 & 29.0 \\
Atlantic & Disabled & \name{}      & 206.2 & 61.3 & 33.6 & 33.6 \\
\midrule
Atlantic & Enabled  & Great Circle & 253.3 & 133.9 & --- & 18.4 \\
Atlantic & Enabled  & CMA-ES       & 159.9 & 59.1 & 36.9 & 48.5 \\
Atlantic & Enabled  & \name{}      & 139.5 & 52.7 & 44.9 & 55.0 \\
\midrule
Pacific  & Disabled & Great Circle & 251.7 & 84.7 & --- & --- \\
Pacific  & Disabled & CMA-ES       & 200.2 & 22.5 & 20.5 & 20.5 \\
Pacific  & Disabled & \name{}      & 194.8 & 20.3 & 22.6 & 22.6 \\
\midrule
Pacific  & Enabled  & Great Circle & 158.0 & 72.1 & --- & 37.2 \\
Pacific  & Enabled  & CMA-ES       &  87.3 & 26.3 & 44.8 & 65.3 \\
Pacific  & Enabled  & \name{}      &  64.1 & 31.6 & 59.4 & 74.5 \\
\bottomrule
\end{tabular}
\end{table*}

Route optimization by CMA-ES alone reduces mean voyage energy by 20--45\% relative to the great-circle baseline of the same WPS mode. The full \name{} pipeline extends these gains to 23--59\%. When combined with wind-assisted propulsion, the total energy savings can reach up to 75\% against a conventional propulsion great-circle route. The largest absolute benefit is on the Trans-Pacific corridor with WPS enabled, where \name{} cuts mean energy from 158.0~MWh (GC) to 64.1~MWh, a reduction of 59.4\%. Without WPS on the same corridor, the optimizer reduces aerodynamic drag alongside hull-resistance and wave-added resistance, but cannot harvest wind thrust, yielding a more modest 22.6\% saving.

\subsubsection{Paired Per-Departure Improvements}
\label{sec:rw-paired}

\begin{figure}[htbp]
    \centering
    \includegraphics[width=\linewidth]{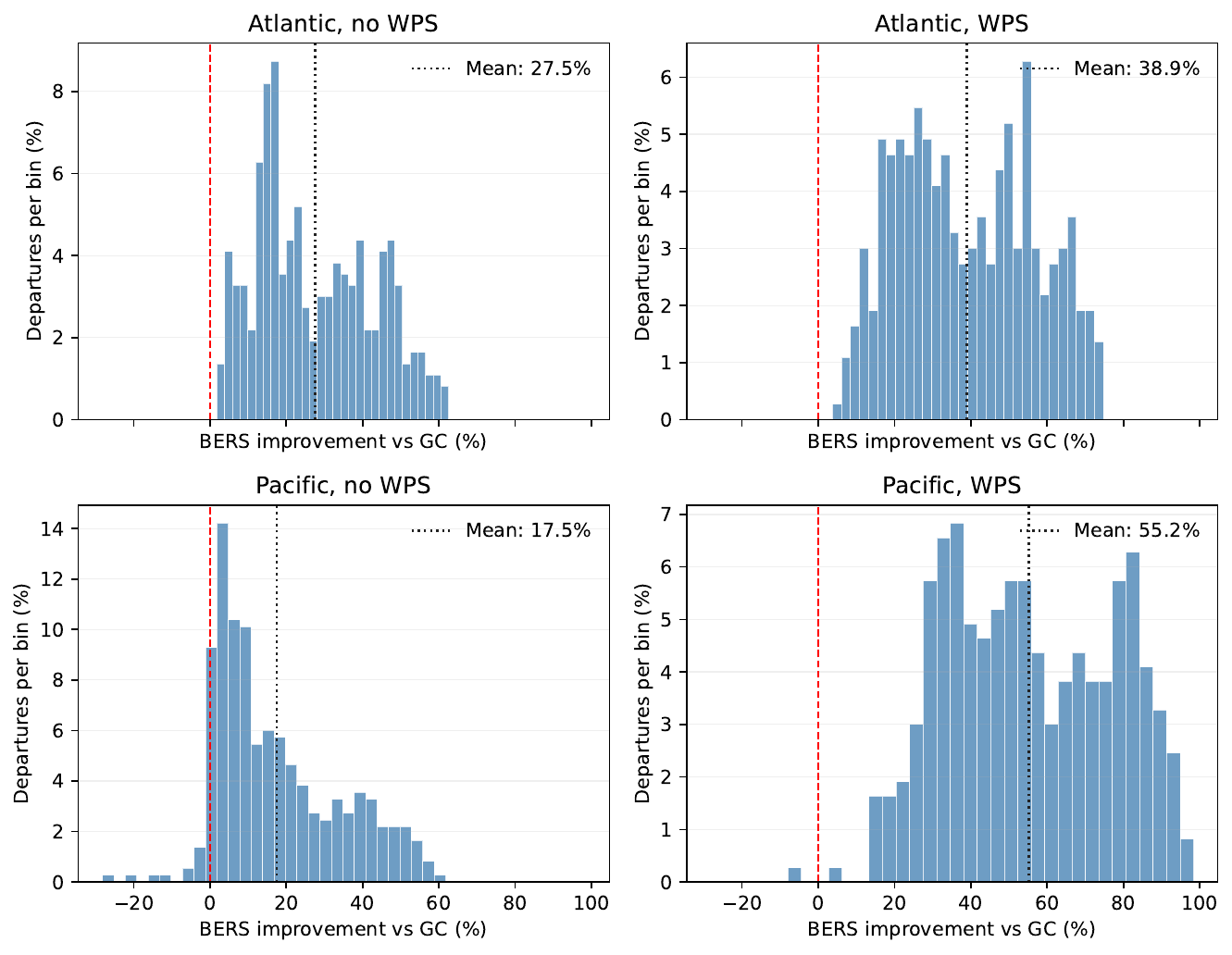}
    \caption{Binned frequency distributions of paired per-departure energy improvement of the full \name{} pipeline over the great-circle baseline of the same propulsion mode, across the 366 departures of 2024. Each histogram uses 30 bins; the dashed line marks zero improvement and the dotted line marks the mean.}
    \label{fig:rw-paired}
\end{figure}

Mean savings over a year of departures do not reveal whether the advantage is systematic or driven by a favourable subset of days. Since all strategies are evaluated on the same 366 departure dates under the same forcing, the comparison is naturally paired, and we therefore report the distribution of per-departure improvements rather than only the difference of means. Table~\ref{tab:rw-paired} summarizes these paired distributions and Figure~\ref{fig:rw-paired} shows their shape.

\begin{DIFnomarkup}

\begin{table*}[htbp]
\centering
\caption{Distribution of paired per-departure improvements over the 366 departures of 2024. Left block: full \name{} against the great-circle baseline of the same propulsion mode. Right block: full \name{} against the CMA-ES-only solution, i.e.\ the marginal contribution of FMS refinement. All values are percentages of the comparator's energy.}
\label{tab:rw-paired}
\begin{tabular}{lrrrrrrrrrr}
\toprule
& \multicolumn{6}{c}{\textbf{\name{} vs great circle}} & \multicolumn{4}{c}{\textbf{\name{} vs CMA-ES only}} \\
\cmidrule(lr){2-7} \cmidrule(lr){8-11}
\textbf{Configuration} & \textbf{Mean} & \textbf{Std} & \textbf{Median} & \textbf{P5} & \textbf{P95} & \textbf{Min} & \textbf{Mean} & \textbf{Std} & \textbf{P5} & \textbf{P95} \\
\midrule
Atlantic, no WPS & 27.5 & 15.3 & 24.0 &  5.7 & 53.8 &  +1.8 &  6.0 &  5.1 & 1.0 & 14.6 \\
Atlantic, WPS    & 38.9 & 17.7 & 37.2 & 12.8 & 67.7 &  +3.7 & 12.5 & 10.1 & 3.3 & 28.7 \\
Pacific, no WPS  & 17.5 & 16.3 & 12.9 &  0.3 & 49.1 & $-28.2$ &  2.6 &  2.1 & 0.8 &  6.0 \\
Pacific, WPS     & 55.2 & 21.8 & 53.1 & 23.2 & 89.1 &  $-8.0$ & 30.6 & 19.7 & 7.0 & 68.1 \\
\bottomrule
\end{tabular}%
\end{table*}
\end{DIFnomarkup}

The advantage is systematic rather than aggregate. In both Atlantic configurations every departure improves on the great circle. The fifth-percentile gain is 5.7\% without WPS and 12.8\% with WPS. On the Pacific, the fifth percentile remains positive, but the tail crosses zero. Fifteen of 366 departures without WPS and one with WPS consume more propulsive energy than the great circle. The worst difference is 28.2\%.

This does not mean that the optimizer preferred a worse value of its own objective. The comparison in Table~\ref{tab:rw-paired} reports propulsive energy only. The great-circle route is not optimized against the land and weather penalties, so it has no incentive to avoid threshold violations. By contrast, \name{} minimizes the penalized objective and may accept a small energy increase to reduce exposure.

Ten of the sixteen exceptions occur in December. Seven fall in the 3--8~December storm sequence in the North Pacific, when the great-circle track is severely exposed. On 6~December, for example, the great circle exceeds the wind threshold by 8.6~m/s and the wave threshold by 6.2~m. It spends 78~h above the wave limit. The \name{} route reduces those exceedances to 0.005~m/s and 0.10~m, respectively. Against propulsive energy alone this trade appears as a loss; against the objective being optimized, it is the intended behaviour.

The right-hand block isolates the contribution of the variational stage. FMS refinement improves on the CMA-ES solution by 2.6--30.6\% on average, with a positive fifth percentile in every configuration, so the refinement is beneficial on at least 95\% of individual departures. It is not universally beneficial: the counts of negative improvements are 4/366 (Atlantic, no WPS; one tie), 4/366 (Atlantic, WPS), 2/366 (Pacific, no WPS; one tie) and 1/366 (Pacific, WPS), with minima of $-24.5\%$, $-61.3\%$, $-0.47\%$ and $-9.6\%$, respectively. Its mean value is largest on the Pacific with WPS (30.6\%) and smallest on the Pacific without WPS (2.6\%).


\subsection{Local Refinement and Criterion~\ref{cr:local-opt}}
\label{sec:rw-local-opt}

The paired analysis provides the direct empirical assessment of Criterion~\ref{cr:local-opt}. FMS improves the CMA-ES objective by 2.6--30.6\% on average and is beneficial on at least 95\% of departures in every configuration (Table~\ref{tab:rw-paired}). It is not universally beneficial, as the small negative tails reported above show, but the result demonstrates systematic local improvement across both corridors and propulsion modes.

Because FMS stops on a stagnation criterion (Section~\ref{sec:algorithm-fms}), these improvements do not by themselves certify that the exported routes are local minima, and we do not claim local optimality for them. Criterion~\ref{cr:local-opt} is addressed through an explicit variational-refinement mechanism and the measurable objective reductions reported above.


\subsection{Effect of Wind Propulsion System}  \label{sec:rw-wps}

Comparing the with/without-WPS columns in Table~\ref{tab:rw-summary} shows the effect of wind assistance for the same routing strategy. On the great-circle route, WPS reduces mean energy by 57~MWh (18.4\%) on the Atlantic and 94~MWh (37.2\%) on the Pacific. When combined with \name{} route optimization, the WPS benefit grows to 67~MWh (32.3\%) and 131~MWh (67.1\%), respectively, because, beyond the aerodynamic drag reduction and wave avoidance available regardless of WPS mode, the optimized route also actively seeks apparent wind angles that maximize wingsail thrust. Combining \name{} routing with WPS reduces mean voyage energy by 55.0\% (Atlantic) and 74.5\% (Pacific) compared with a propulsion-only great-circle voyage.



\subsection{Seasonal Variability}  \label{sec:rw-seasonal}

Figure~\ref{fig:rw-routes} provides representative monthly route geometries that help interpret seasonal patterns. The great-circle baseline exhibits pronounced seasonal oscillations driven by the intensification and weakening of the mid-latitude westerlies, with peaks in winter and troughs in summer. The optimized routes substantially damp this variability: \name{} achieves its largest absolute savings in winter, when strong winds simultaneously increase propulsive demand on the great-circle route and provide the greatest sail thrust on optimized routes, and its smallest savings in summer.

\begin{figure}[htbp]
\centering
\begin{subfigure}[t]{0.45\linewidth}
    \centering
    \includegraphics[width=\linewidth]{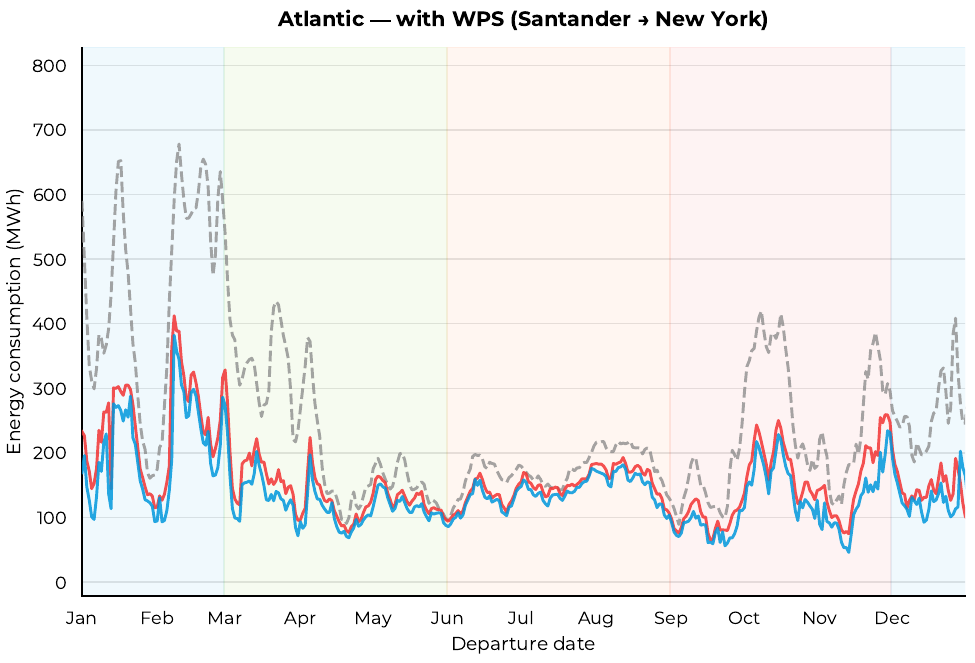}
    \caption{Trans-Atlantic, WPS enabled.}
\end{subfigure}
\begin{subfigure}[t]{0.45\linewidth}
    \centering
    \includegraphics[width=\linewidth]{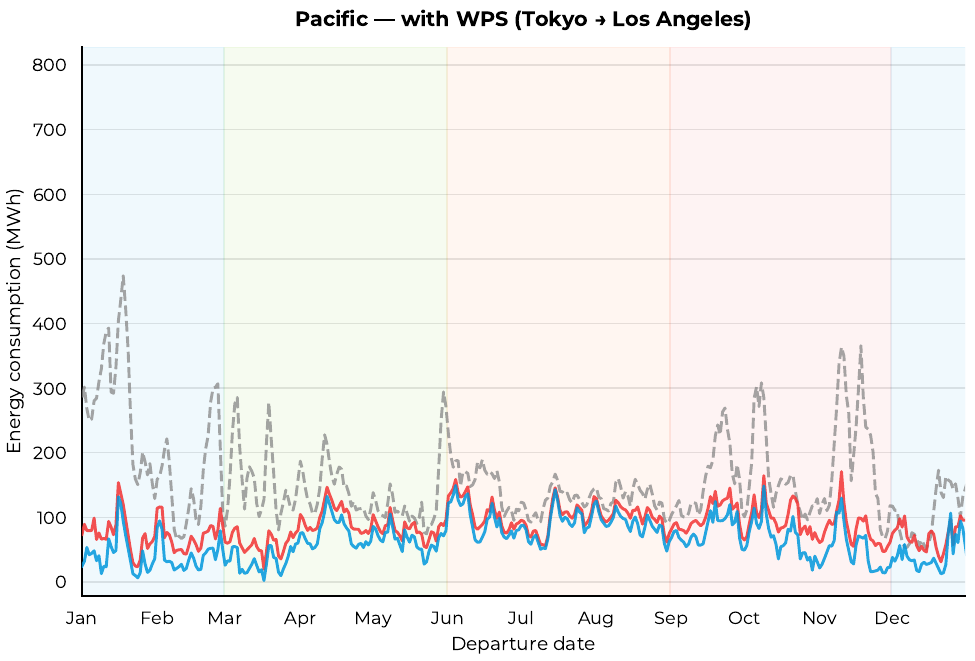}
    \caption{Trans-Pacific, WPS enabled.}
\end{subfigure}
\begin{subfigure}[t]{0.45\linewidth}
    \centering
    \includegraphics[width=\linewidth]{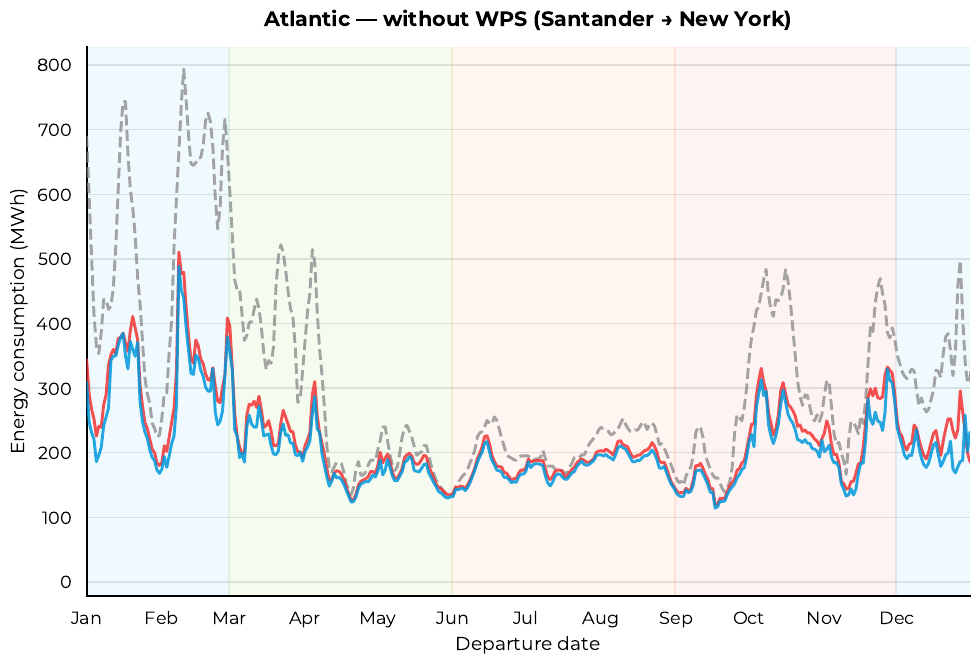}
    \caption{Trans-Atlantic, WPS disabled.}
\end{subfigure}
\begin{subfigure}[t]{0.45\linewidth}
    \centering
    \includegraphics[width=\linewidth]{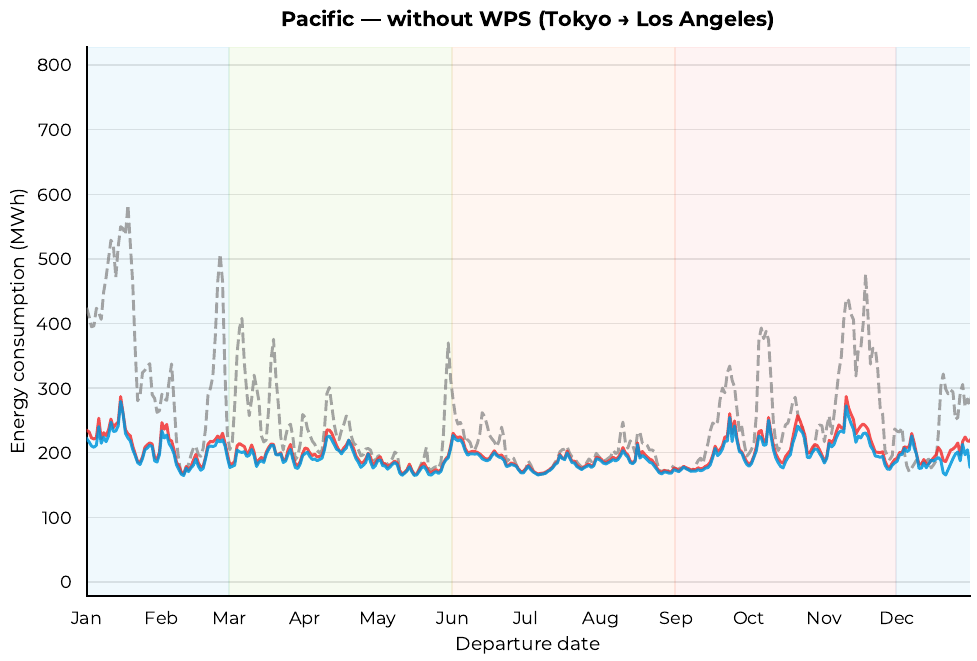}
    \caption{Trans-Pacific, WPS disabled.}
\end{subfigure}
\caption{Daily voyage energy throughout 2024 for the great-circle baseline (dashed grey), CMA-ES alone (red), and full \name{} (blue).}
\label{fig:rw-seasonality}
\end{figure}

Table~\ref{tab:rw-seasonal} summarizes seasonal savings of the full \name{} pipeline with WPS relative to the WPS great-circle baseline. Winter delivers the highest gains: 53.8\% for the Atlantic and 77.0\% for the Pacific, reflecting the intense and consistent westerly circulation during boreal winter. Summer savings remain appreciable (20.8\% and 33.1\%), demonstrating that route optimization provides meaningful benefits throughout the year.

\begin{table}[htbp]
\centering
\caption{Seasonal energy savings (\%) of the full \name{} pipeline with WPS enabled versus the great-circle baseline with WPS.}
\label{tab:rw-seasonal}
\begin{tabular}{lrr}
\toprule
\textbf{Season} & \textbf{Atlantic (\%)} & \textbf{Pacific (\%)} \\
\midrule
Winter & 53.8 & 77.0 \\
Spring & 46.1 & 58.9 \\
Summer & 20.8 & 33.1 \\
Autumn & 47.6 & 62.2 \\
\bottomrule
\end{tabular}
\end{table}


\subsection{Weather-Threshold Exposure}  \label{sec:rw-violations}

\begin{figure*}[htbp]
    \centering
    \includegraphics[width=\textwidth]{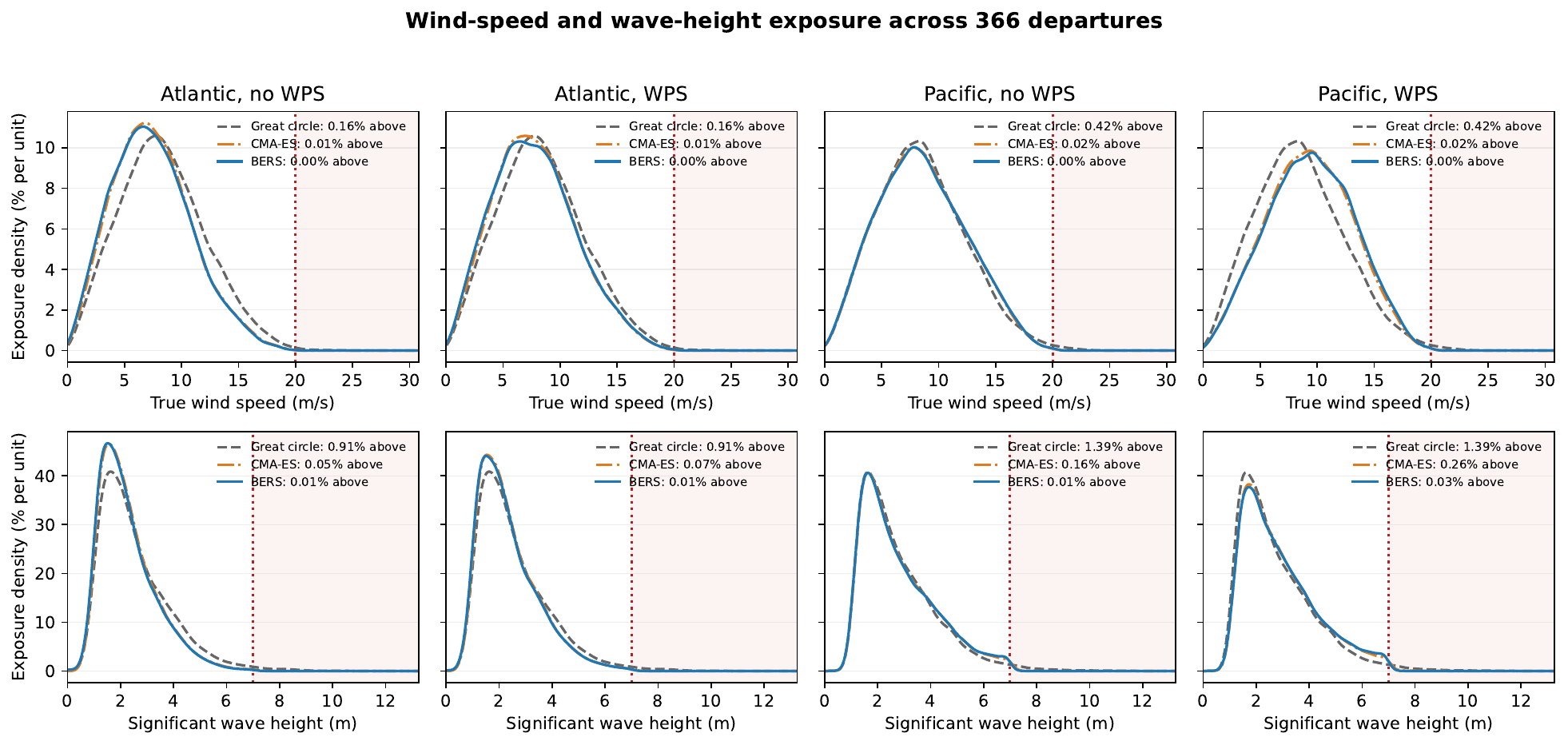}
    \caption{Time-weighted segment-level distributions over all 366 departures for the great-circle baseline, CMA-ES route, and final \name{} route. The upper panels show true wind speed and the lower panels significant wave height for each corridor and propulsion configuration. The red dotted lines mark the benchmark thresholds of 20~m/s and 7~m; the percentage beside each method in the legend is the share of its total evaluated route time above the corresponding threshold. The continuous curves are Gaussian-smoothed, duration-weighted densities evaluated on a common fine grid within each environmental variable.}
    \label{fig:rw-violation-curves}
\end{figure*}

The wind-speed and wave-height thresholds of Section~\ref{sec:land-avoidance} are imposed as soft rather than hard constraints, so it is necessary to report how often, by how much, and for how long the optimized routes cross them. Figure~\ref{fig:rw-violation-curves} compares the time-weighted distributions of true wind speed and significant wave height along the great-circle, CMA-ES, and final \name{} routes for every configuration. Table~\ref{tab:rw-violations} then gives detailed \name{}--great-circle statistics at the departure level. We count a departure as exceeding a threshold if any evaluated segment does, and report the mean and maximum of the per-departure peak exceedance together with the mean and maximum time spent above the threshold. Before midpoint evaluation, route longitudes are unwrapped at the antimeridian, converted to the longitude convention of the corresponding ERA5 grid, and checked against that grid's bounds.

The distributions show the progression from an unoptimized geometric baseline to global search and then variational refinement. Depending on the corridor, the great-circle route spends 0.16--0.42\% of evaluated time above the wind threshold and 0.91--1.39\% above the wave threshold. CMA-ES reduces those ranges to 0.01--0.02\% and 0.05--0.26\%, respectively; after FMS, the corresponding \name{} exposure is 0.00\% for wind and 0.01--0.03\% for waves. These time-weighted percentages complement, rather than replace, the route-level incidence and exceedance magnitudes in Table~\ref{tab:rw-violations}.

\begin{DIFnomarkup}
\begin{table*}[htbp]
\centering
\caption{Exposure to the wind-speed (20~m/s) and wave-height (7~m) thresholds over the 366 departures of 2024. ``Departures'' is the percentage of departures whose route crosses the threshold at least once. The remaining columns are conditional on those departures: ``peak exceedance'' is the largest amount by which a departure exceeds the threshold, averaged over the violating departures and at its worst, and ``time above'' is the duration spent beyond the threshold on the same basis. Exceedances are in m/s for wind and m for waves.}
\label{tab:rw-violations}
\begin{tabular}{llrrrrr}
\toprule
& & & \multicolumn{2}{c}{\textbf{Peak exceedance}} & \multicolumn{2}{c}{\textbf{Time above (h)}} \\
\cmidrule(lr){4-5} \cmidrule(lr){6-7}
\textbf{Configuration} & \textbf{Threshold} & \textbf{Departures (\%)} & \textbf{Mean} & \textbf{Max} & \textbf{Mean} & \textbf{Max} \\
\midrule
Atlantic, no WPS --- GC     & Wind & 10.9 & 1.423 & 6.57 &  5.3 & 18.0 \\
Atlantic, no WPS --- \name{} & Wind &  0.0 & 0.000 & 0.00 &  0.0 &  0.0 \\
Atlantic, no WPS --- GC     & Wave & 18.0 & 1.423 & 4.99 & 17.8 & 46.0 \\
Atlantic, no WPS --- \name{} & Wave &  0.8 & 0.056 & 0.11 &  5.3 & 12.0 \\
\midrule
Atlantic, WPS --- GC     & Wind & 10.9 & 1.423 & 6.57 &  5.3 & 18.0 \\
Atlantic, WPS --- \name{} & Wind &  0.3 & 0.009 & 0.01 &  2.0 &  2.0 \\
Atlantic, WPS --- GC     & Wave & 18.0 & 1.423 & 4.99 & 17.8 & 46.0 \\
Atlantic, WPS --- \name{} & Wave &  1.1 & 0.071 & 0.13 &  2.5 &  4.0 \\
\midrule
Pacific, no WPS --- GC     & Wind & 24.3 & 2.016 & 10.64 & 10.0 & 45.9 \\
Pacific, no WPS --- \name{} & Wind &  0.6 & 0.140 &  0.19 &  3.0 &  4.0 \\
Pacific, no WPS --- GC     & Wave & 33.3 & 1.539 &  6.17 & 24.3 & 85.9 \\
Pacific, no WPS --- \name{} & Wave &  1.4 & 0.015 &  0.03 &  2.8 &  4.0 \\
\midrule
Pacific, WPS --- GC     & Wind & 24.3 & 2.016 & 10.64 & 10.0 & 45.9 \\
Pacific, WPS --- \name{} & Wind &  0.6 & 0.247 &  0.32 &  5.0 &  6.0 \\
Pacific, WPS --- GC     & Wave & 33.3 & 1.539 &  6.17 & 24.3 & 85.9 \\
Pacific, WPS --- \name{} & Wave &  1.1 & 0.564 &  2.16 & 14.5 & 47.9 \\
\bottomrule
\end{tabular}%
\end{table*}
\end{DIFnomarkup}

Both incidence and severity fall sharply on the final routes. Atlantic wind incidence falls from 10.9\% on the great circle to 0--0.3\%, and wave incidence from 18.0\% to 0.8--1.1\%. Pacific wind and wave incidence fall from 24.3\% and 33.3\% to at most 0.6\% and 1.4\%, respectively. Across the nonzero cells, conditional mean peak exceedance falls by factors of approximately 2.7--155, while conditional mean duration falls by factors of 1.7--8.7. The largest optimized-route exceedance is 0.32~m/s for wind and 2.16~m for waves; the great-circle maxima are 10.64~m/s and 6.17~m. The largest remaining wave event is confined to the Pacific WPS case: only four departures exceed 7~m, although their conditional mean duration is 14.5~h.

This is consistent with the squared-excess penalty of Equation~\eqref{eq:penalization-env}. In most configurations, the optimized routes turn the great circle's long, deep excursions into rare marginal crossings. The Pacific WPS case is the exception. Its four remaining wave exceedances are neither marginal nor brief. A soft penalty can accept such an excursion when avoiding it costs too much propulsive energy.

The weights represent one point on an energy--exposure trade-off, not an optimum proved by this study. A strict operator limit would require a hard feasibility rule, a departure-time decision, or both. More generally, future work should estimate the Pareto frontier instead of fixing one weighted sum.


\section{Discussion} \label{sec:discussion}

The results of this study show that \name{} effectively balances exploration, constraint handling, and solution refinement in the context of maritime weather routing. By combining the global search capabilities of the Covariance Matrix Adaptation Evolution Strategy (CMA-ES) with the local refinement of the Ferraro-Mart\'in de Diego-Sato (FMS) algorithm, \name{} can be evaluated systematically against the proposed routing criteria in synthetic tests and real-ocean benchmark scenarios. It produces continuous, obstacle-aware routes, remains adaptable to changing conditions, and supports a range of cost functions.

The results in Section~\ref{sec:synthetic-vectorfields} show that, within the benchmark settings, \name{} addresses the seven design goals introduced earlier. The continuous B\'ezier representation avoids grid artifacts; the land penalty and FMS rollback reduce sampled violations near obstacles; and the two cost formulations cover both time-optimal and fixed-arrival objectives while maintaining constant engine load. The variational stage further seeks a low-cost extremal from the CMA-ES solution. These controlled settings emulate many practical challenges, but do not establish geometric feasibility or local minimality in every case and do not encompass the full variability of real-world routing.

A natural question is whether \name{} can support dynamic re-planning during a voyage, as operational weather routing requires periodic route updates based on revised forecasts. In the SWOPP3 benchmark, the reported optimization time was under 0.01~CPU-hours per departure, compared with more than 10~CPU-hours for some competing entries. This batch result suggests that the optimizer can fit within a 6--12~hour update interval, but it is not a closed-loop demonstration: operational latency would also include forecast acquisition, data preparation, quality control and route approval. The previous solution could seed CMA-ES and provide the initial curve for FMS in a rolling-horizon controller. Testing that warm-started model-predictive-control workflow with ensemble forecasts and quantified forecast error remains future work, so we do not claim robustness to forecast uncertainty from the reanalysis experiments reported here.

From a computational perspective, \name{} maintains a practical trade-off between quality and runtime on the hardware used (NVIDIA RTX A6000 GPU). CMA-ES typically converges within seconds. The parameter studies indicate the expected trade-off between expressiveness ($K$) and optimization difficulty, and a context-dependent role for the initial search variance $\sigma_0$. FMS adds cost, especially near land or under complex flow, but is not always required. Selectively applying FMS when CMA-ES fails to produce a well-refined result could significantly reduce the computation time. This strategy could be guided by adaptive stopping criteria or confidence thresholds.

The real-ocean experiments show several consistent trends. The Trans-Pacific route with WPS gives the largest gains: 59\% from routing alone and 75\% when routing and wingsails are combined. Winter gives the highest savings on both corridors, while summer savings remain 21--33\%. The WPS saving grows from 18--37\% on great-circle routes to 32--67\% on \name{} routes.

FMS improves on CMA-ES in at least 95\% of departures in every configuration, with mean gains of 2.6--30.6\% (Table~\ref{tab:rw-paired}). The result is systematic local improvement; local optimality of the exported routes is not certified.

Weather exposure also falls sharply. Relative to the great circle, the conditional mean threshold exceedance falls by factors of approximately 2.7--155. Its conditional mean duration falls by factors of 1.7--8.7 (Table~\ref{tab:rw-violations}). These results support the chosen scalarization in the tested setting, but they are not a formal sensitivity analysis or a general feasibility guarantee.

The scope remains limited to one vessel, two corridors, and ERA5 reanalysis. Reanalysis represents weather hindsight rather than a forecast, and the performance model represents one hull and sail configuration. The savings should not be read as an operational guarantee or transferred directly to other vessels. The larger savings than the 0--7\% often reported for conventional weather routing \cite{imo_greenvoyage2050_weatherrouting, sofar2024} are specific to these wind-assisted, fixed-time scenarios.

These results also reveal a structural limitation of the present study: the experiments cover two specific open-ocean corridors for a single vessel type. Although the corridors are representative of major trans-oceanic trade lanes and the 366-departure design captures a full year of ERA5 variability, the magnitude of savings will depend on vessel dimensions, sail area, propulsion power, and the prevailing wind climatology of the chosen route. Narrow waterways such as the Suez or Panama canals remain challenging for a smooth, low-dimensional trajectory representation: a single continuous B\'ezier curve with a limited number of control points cannot reliably represent tight, elongated passages without violating feasibility constraints. A hybrid planner that switches to a corridor-constrained method for narrow sections and reconnects to the global B\'ezier parametrization in open waters would likely address this limitation.


\section{Concluding Remarks and Future Directions} \label{sec:conclusion}

This paper presented \name{}, a two-phase optimization strategy for maritime weather routing that integrates global exploration using CMA-ES with local trajectory refinement through FMS. The method is designed to address key challenges in trajectory planning, including flow adaptation, obstacle avoidance, and scheduling constraints.

Through synthetic benchmarks, \name{} demonstrated consistent performance in a diverse set of scenarios, including steady-state and time-dependent vector fields, fixed-speed travel, and just-in-time arrival. The algorithm reproduced or improved known objective values from the literature and reduced sampled land violations under varying land complexities.

Synthetic results alone do not establish general applicability, so we also assessed \name{} on the real-ocean SWOPP3 scenarios. The study covers two trans-oceanic corridors and 366 departures in 2024, with ERA5 reanalysis and the reconstructed energy model in \ref{sec:appendix-performance-model}.

Route optimization reduces mean propulsive energy by 23--59\% relative to the corresponding great-circle baseline. Combining route optimization with wingsails reaches 75\% relative to a propulsion-only great circle. FMS improves on CMA-ES on at least 95\% of departures in every configuration. This is evidence of systematic local improvement; local optimality of the exported routes is not certified.

These results apply to the tested vessel, corridors, and reanalysis year. The speed tails and remaining weather-threshold crossings limit their operational interpretation. The final polygon audit finds no land intersections.

Future work will focus on operational deployment and methodological extensions. An adaptive application of FMS, triggered when CMA-ES results lack smoothness or feasibility, could reduce runtime without degrading quality. A coupled Newton polishing step based on the complete discrete Hessian could help establish certified local minima. The energy--weather trade-off should be treated explicitly as a multi-objective problem, producing a Pareto frontier from which an operator can choose. For narrow passages, a hybrid planner that switches to a corridor-constrained graph or sampling-based method would likely be more reliable than a single global parametrization. We also plan to incorporate more realistic cost functions derived from naval engineering models and weather forecasts.

In summary, \name{} offers a flexible and modular approach to weather routing, combining global evolutionary search with FMS variational refinement. The real-world experiments demonstrate substantial energy savings for wind-assisted vessels and systematic local improvement from FMS, while Criterion~\ref{cr:local-opt} remains an objective rather than a guarantee. Within the same SWOPP3 benchmark comparison considered in this paper, \name{} is competitive not only in solution quality but also in operational efficiency, ranking within 1.1\% of the best-performing submission while achieving the lowest reported computation time by more than one order of magnitude. Further empirical validation on additional vessel types and corridors, and methodological extensions for narrow-waterway routing, remain as future work.


\section*{Acknowledgements}  \label{sec:acknowledgments}

This work was carried out within the framework of the BBVA-Funding project {``Optimization of maritime routes for a more efficient, safer, and decarbonized maritime transport''}. This paper was partially funded by MICIU / AIE /10.13039/501100011033 / FEDER, UE, Grants No. PID2024-155187OB-I00 (KAN4MET) and PID2021-122154NB-I00 (OP4ML), co-financed by the European Regional Development Fund, ``ERDF - A Way of Making Europe''.


\bibliography{references}

@article{era5,
  author    = {Hersbach, Hans and Bell, Bill and Berrisford, Paul and Hirahara, Shoji and Hor{\'a}nyi, Andr{\'a}s and Mu{\~n}oz-Sabater, Joaqu{\'i}n and Nicolas, Julien and Peubey, Carole and Radu, Raluca and Schepers, Dinand and others},
  title     = {{ERA5} global reanalysis},
  journal   = {Quarterly Journal of the Royal Meteorological Society},
  year      = {2020},
  volume    = {146},
  number    = {730},
  pages     = {1999--2049},
  doi       = {10.1002/qj.3803}
}

@article{Hansen2003,
  author={Hansen, Nikolaus and M\"{u}ller, Sibylle D. and Koumoutsakos, Petros},
  journal={Evolutionary Computation}, 
  title={Reducing the Time Complexity of the Derandomized Evolution Strategy with Covariance Matrix Adaptation ({CMA-ES})}, 
  year={2003},
  volume={11},
  number={1},
  pages={1--18},
  doi={10.1162/106365603321828970}}

@article{Boehm1999,
title = {On de {C}asteljau's algorithm},
journal = {Computer Aided Geometric Design},
volume = {16},
number = {7},
pages = {587--605},
year = {1999},
issn = {0167-8396},
doi = {https://doi.org/10.1016/S0167-8396(99)00023-0},
author = {Wolfgang Boehm and Andreas M\"{u}ller},
}

@Article{Ferraro2021,
  author    = {Sebasti\'an J. Ferraro and David Mart\'in {de Diego} and Rodrigo T. Sato Mart\'in de Almagro},
  journal   = {IFAC-PapersOnLine},
  title     = {Parallel iterative methods for variational integration applied to navigation problems},
  year      = {2021},
  issn      = {2405-8963},
  month     = jul,
  pages     = {321--326},
  volume    = {54},
  doi       = {10.1016/J.IFACOL.2021.11.097},
  publisher = {Elsevier},
}

@article{Ferraro2022,
	title        = {A parallel iterative method for variational integration},
	author       = {Ferraro, Sebasti{\'a}n J and Mart{\'i}n de Diego, David and Sato Mart{\'i}n de Almagro, Rodrigo Takuro},
	year         = 2025,
	journal      = {Journal of Computational Dynamics},
	volume       = 12,
	number       = 4,
	pages        = {528--553},
	doi          = {10.3934/jcd.2025006}
}

@article{Precioso2024,
    title={Hybrid Search method for {Z}ermelo's navigation problem}, 
    author={Daniel Precioso and Robert Milson and Louis Bu and Yvonne Menchions and David Gómez-Ullate},
    year={2024},
    journal={Computational and Applied Mathematics},
    volume={43},
    number={4},
    pages={250},
    doi={10.1007/s40314-024-02756-w}
}

@phdthesis{Precioso2023Thesis,
    title={Applications of machine learning and data science to the blue economy sustainable fishing and weather routing},
    author={Precioso Garcel{\'a}n, Daniel},
    year={2023},
    school={Universidad de Cádiz}
}

@article{Jimenez2024,
    title={{HADAD}: Hexagonal {A}-Star with Differential Algorithm Designed for weather routing}, 
    author={Javier Jiménez-de-la-Jara and Daniel Precioso and Louis Bu and Victoria Redondo-Neble and Robert Milson and Rafael Ballester-Ripoll and David Gómez-Ullate},
    year={2025},
    journal={Ocean Engineering},
    volume={319},
    pages={120050},
    doi={10.1016/j.oceaneng.2024.120050}
}

@article{mannarini2024visir,
	title        = {{VISIR-2: ship weather routing in Python}},
	author       = {Mannarini, Gianandrea and Salinas, Mario Leonardo and Carelli, Lorenzo and Petacco, Nicola and Orovi{\'c}, Josip},
	year         = 2024,
	journal      = {Geoscientific Model Development},
	publisher    = {Copernicus Publications G{\"o}ttingen, Germany},
	volume       = 17,
	number       = 10,
	pages        = {4355--4382}
}

@article{Charalambopoulos2023,
	title        = {Efficient ship weather routing using probabilistic roadmaps},
	author       = {Nikolaos Charalambopoulos and Elias Xidias and Andreas Nearchou},
	year         = 2023,
	journal      = {Ocean Engineering},
	volume       = 273,
	pages        = 114031,
	doi          = {https://doi.org/10.1016/j.oceaneng.2023.114031},
	issn         = {0029-8018}
}

@article{szlapczynski2023ship,
  title={{Ship weather routing featuring w-MOEA/D and uncertainty handling}},
  author={Szlapczynski, Rafal and Szlapczynska, Joanna and Vettor, Roberto},
  journal={Applied Soft Computing},
  volume={138},
  pages={110142},
  year={2023},
  publisher={Elsevier}
}

@article{Grandcolas2022,
	title        = {A Metaheuristic Algorithm for Ship Weather Routing},
	author       = {Grandcolas, St{\'e}phane},
	year         = 2022,
	journal = {SN Operations Research Forum},
	booktitle    = {Operations Research Forum},
	volume       = 3,
	pages        = 35,
	organization = {Springer}
}

@article{Grifoll2022,
	title        = {{A comprehensive ship weather routing system using CMEMS products and A* algorithm}},
	author       = {Grifoll, Manel and Bor{\'e}n, Clara and Castells-Sanabra, Marcella},
	year         = 2022,
	journal      = {Ocean Engineering},
	publisher    = {Elsevier},
	volume       = 255,
	pages        = 111427
}

@article{Zhao2022,
	title        = {Multi-objective weather routing algorithm for ships based on hybrid particle swarm optimization},
	author       = {Zhao, Wei and Wang, Hongbo and Geng, Jianning and Hu, Wenmei and Zhang, Zhanshuo and Zhang, Guangyu},
	year         = 2022,
	journal      = {Journal of Ocean University of China},
	publisher    = {Springer},
	volume       = 21,
	number       = 1,
	pages        = {28--38}
}

@article{Techy2011,
  author       = {Laszlo Techy},
  title        = {Optimal navigation in planar time-varying flow: {Z}ermelo's problem revisited},
  journal      = {Intelligent Service Robotics},
  year         = {2011},
  volume       = {4},
  number       = {4},
  pages        = {271--283},
  month        = {October},
  issn         = {1861-2784},
  doi          = {10.1007/s11370-011-0092-9},
  url          = {https://doi.org/10.1007/s11370-011-0092-9}
}

@article{mannarini2019visir,
  title={{VISIR-1. b: Ocean surface gravity waves and currents for energy-efficient navigation}},
  author={Mannarini, Gianandrea and Carelli, Lorenzo},
  journal={Geoscientific Model Development},
  volume={12},
  number={8},
  pages={3449--3480},
  year={2019},
  publisher={Copernicus GmbH}
}

@article{Shadden2005,
title = {Definition and properties of Lagrangian coherent structures from finite-time Lyapunov exponents in two-dimensional aperiodic flows},
journal = {Physica D: Nonlinear Phenomena},
volume = {212},
number = {3},
pages = {271-304},
year = {2005},
issn = {0167-2789},
doi = {https://doi.org/10.1016/j.physd.2005.10.007},
url = {https://www.sciencedirect.com/science/article/pii/S0167278905004446},
author = {Shawn C. Shadden and Francois Lekien and Jerrold E. Marsden},
}

@article{Gunnarson2021,
  title={Learning efficient navigation in vortical flow fields},
  author={Gunnarson, Peter and Mandralis, Ioannis and Novati, Guido and Koumoutsakos, Petros and Dabiri, John O},
  journal={Nature communications},
  volume={12},
  number={1},
  pages={7143},
  year={2021},
  publisher={Nature Publishing Group UK London}
}

@article{Zermelo1931,
author = {Zermelo, E.},
title = {Über das Navigationsproblem bei ruhender oder veränderlicher Windverteilung},
journal = {ZAMM - Journal of Applied Mathematics and Mechanic},
volume = {11},
number = {2},
pages = {114-124},
doi = {https://doi.org/10.1002/zamm.19310110205},
eprint = {https://onlinelibrary.wiley.com/doi/pdf/10.1002/zamm.19310110205},
year = {1931}
}

@article{bezier1972numerical,
  title={Numerical control-mathematics and applications},
  author={B{\'e}zier, Pierre},
  journal={Translated by AR Forrest},
  year={1972},
}

@misc{decasteljau1985mathematiques,
  title={Math{\'e}matiques et CAO. Volume 2: formes {\`a} p{\^o}les},
  author={De Casteljau, Paul},
  year={1985}
}

@book{farin2001curves,
  title={Curves and surfaces for CAGD: a practical guide},
  author={Farin, Gerald},
  year={2001},
  publisher={Elsevier}
}

@article{hansen2001completely,
  title={Completely derandomized self-adaptation in evolution strategies},
  author={Hansen, Nikolaus and Ostermeier, Andreas},
  journal={Evolutionary computation},
  volume={9},
  number={2},
  pages={159--195},
  year={2001},
  publisher={MIT Press}
}

@article{hansen2016cma,
  title={The CMA evolution strategy: A tutorial},
  author={Hansen, Nikolaus},
  journal={arXiv preprint arXiv:1604.00772},
  year={2016}
}

@misc{hansen2019pycma,
  author       = {Nikolaus Hansen and Youhei Akimoto and Petr Baudis},
  title        = {{CMA-ES/pycma} on {G}ithub},
  howpublished = {Zenodo, DOI:10.5281/zenodo.2559634},
  month        = feb,
  year         = 2019,
  doi          = {10.5281/zenodo.2559634},
  url          = {https://doi.org/10.5281/zenodo.2559634},
}

@article{perlin1985image,
  title={An image synthesizer},
  author={Perlin, Ken},
  journal={ACM Siggraph Computer Graphics},
  volume={19},
  number={3},
  pages={287--296},
  year={1985},
  publisher={ACM New York, NY, USA}
}

@inproceedings{lagae2010survey,
  title={A survey of procedural noise functions},
  author={Lagae, Ares and Lefebvre, Sylvain and Cook, Rob and DeRose, Tony and Drettakis, George and Ebert, David S and Lewis, John P and Perlin, Ken and Zwicker, Matthias},
  booktitle={Computer Graphics Forum},
  volume={29},
  pages={2579--2600},
  year={2010},
  organization={Wiley Online Library}
}

@misc{jax2018github,
  author = {James Bradbury and Roy Frostig and Peter Hawkins and Matthew James Johnson and Chris Leary and Dougal Maclaurin and George Necula and Adam Paszke and Jake Vander{P}las and Skye Wanderman-{M}ilne and Qiao Zhang},
  title = {{JAX}: composable transformations of {P}ython+{N}um{P}y programs},
  url = {http://github.com/jax-ml/jax},
  version = {0.3.13},
  year = {2018},
}

@misc{imo_greenvoyage2050_weatherrouting,
  author       = {{International Maritime Organization}},
  title        = {GreenVoyage2050: Ship Energy Efficiency Measures -- Weather Routing},
  year         = {2022},
  howpublished = {\url{https://greenvoyage2050.imo.org/}},
  note         = {Accessed: 2025-10-09}
}

@techreport{sofar2024,
  author       = {{Sofar Ocean}},
  title        = {Fuel and Emissions Savings from Real-Time Voyage Optimization},
  institution  = {Sofar Ocean Technologies},
  year         = {2024},
  note         = {Technical Report},
  howpublished = {\url{https://www.sofarocean.com/reports/fuel-and-emissions-savings}},
  urldate      = {2025-10-09}
}

@misc{imo_ghg_factors,
  author       = {{International Maritime Organization}},
  title        = {Fourth IMO GHG Study 2020},
  year         = {2020},
  howpublished = {\url{https://www.imo.org/en/OurWork/Environment/Pages/Fourth-IMO-GHG-Study-2020.aspx}},
  note         = {Greenhouse gas emission factors and analysis for international shipping}
}

@misc{imo_2023_ghg_strategy,
  author       = {{International Maritime Organization}},
  title        = {2023 IMO Strategy on Reduction of GHG Emissions from Ships},
  year         = {2023},
  howpublished = {\href{https://www.imo.org/en/ourwork/environment/pages/2023-imo-strategy-on-reduction-of-ghg-emissions-from-ships.aspx}{IMO website}},
  note         = {Adopted by MEPC 80 (Resolution MEPC.377(80))},
  urldate      = {2025-10-09}
}

@misc{swopp3,
  author       = {{RISE Research Institutes of Sweden}},
  title        = {SWOPP 3 Workshop: Joint industry project on weather routing and speed optimization},
  year         = {2026},
  note         = {Project webpage, accessed June 2026},
  url          = {https://www.ri.se/en/maritime/project/joint-industry-project-on-weather-routing-and-speed-optimization}
}

@techreport{IMO_MSC1228,
  author       = {{International Maritime Organization}},
  title        = {Revised Guidance to the Master for Avoiding Dangerous
                  Situations in Adverse Weather and Sea Conditions},
  institution  = {IMO},
  year         = {2007},
  number       = {MSC.1/Circ.1228},
  note         = {11 January 2007}
}

@techreport{DNVGL_seakeeping,
  author       = {{DNV GL}},
  title        = {Wave Loads},
  institution  = {DNV GL AS},
  year         = {2018},
  number       = {DNVGL-CG-0130},
  type         = {Class Guideline},
  note         = {Edition January 2018}
}

@article{BaoRoblesShen2004,
  author  = {Bao, David and Robles, Colleen and Shen, Zhongmin},
  title   = {Zermelo navigation on Riemannian manifolds},
  journal = {Journal of Differential Geometry},
  volume  = {66},
  number  = {3},
  pages   = {377--435},
  year    = {2004},
  doi     = {10.4310/jdg/1098137838}
}


\newpage

\appendix

\section{Definition of Synthetic Vector Fields}
\label{sec:appendix-synthetic-fields}

\subsection{Circular Vector Field}

Originally formulated by Techy \textit{et al.}~\cite{Techy2011}, the Circular vector field models a rigid-body rotational flow centred at the origin. The field induces uniform circular trajectories across the plane, governed by the angular velocity $\omega$:

\begin{equation}
\label{eq:circular}
\mathbf{w}(x, y) =
\begin{bmatrix}
- \omega \, y \\
\omega \, x
\end{bmatrix},
\end{equation}

\noindent where $\omega \in \mathbb{R}$ controls the rotational direction and speed. Positive values correspond to counter-clockwise flow, while negative values induce clockwise motion. In this benchmark, we set $\omega = -0.9$, resulting in clockwise rotation.

The routing task consists of identifying the optimal-time route from $\mathbf{x}_A = \left( \cos( \pi/6 ), \sin( \pi/6 ) \right)$ to $\mathbf{x}_B = (0, 1)$, under the constraint of constant unit vessel speed. As the vector field imparts a uniform angular velocity, it offers a clean test bed for assessing trajectory planning and control strategies in rotational flow environments.

The assumption of fixed vessel speed provides a controlled test of Criterion~\ref{cr:constant-power}; it should not be interpreted as a propulsion-system certification. These simple conditions make the Circular field a natural starting point for validating weather-routing algorithms. While it lacks environmental complexity, it serves to confirm baseline behaviour under the stated constant-speed assumption. The following vector fields continue to test this criterion while introducing additional challenges, such as spatial nonlinearity and temporal dynamics.

\subsection{Four-Vortices Vector Field}

This vector field, originally introduced by Ferraro \textit{et al.}~\cite{Ferraro2021}, is constructed by superimposing four rotational flow components centred at fixed locations. It is defined by:

\begin{equation}
\begin{split}
\label{eq:four-vortices}
\mathbf{w}&(x, y) =\\
&s \left( -R_{2, 2}(x, y) - R_{4, 4}(x, y) - R_{2, 5}(x, y) + R_{5, 1}(x, y) \right),
\end{split}
\end{equation}

\noindent where each vortex component $R_{a,b}(x, y)$ takes the form:

\begin{equation}
R_{a,b}(x, y) = \frac{1}{3 \left( (x - a)^2 + (y - b)^2 \right) + 1}
\begin{bmatrix}
-(y - b) \\ x - a
\end{bmatrix},
\end{equation}

\noindent where the scalar $s = 1.7$ is selected to limit the maximum flow speed to approximately 1. The optimization goal is to compute a time-optimal trajectory from $\mathbf{x}_A = (0, 0)$ to $\mathbf{x}_B = (6, 2)$, under a fixed-speed constraint of 1 unit. Ferraro \textit{et al.}~\cite{Ferraro2021} reported an optimal travel time of $T = 8.95$ under these conditions.

Each vortex decays with the square of distance from its centre, generating sharp local velocity gradients. These induce non-linearities in the cost landscape, requiring continuous trajectory optimization and sufficiently dense along-path evaluation to capture optimal paths. Otherwise, narrow low-cost channels between vortex cores may be missed between sampled points, motivating criterion~\ref{cr:continuous-space}.

\subsection{Double-Gyre Vector Field}

The Double-Gyre field, introduced by Shadden \textit{et al.}~\cite{Shadden2005}, models a time-periodic, two-dimensional flow that approximates features of large-scale geophysical circulations. It has been further explored in oceanographic studies~\cite{Gunnarson2021}. The system is parametrized by $A$ (amplitude), $\epsilon$ (oscillation strength), and $\omega$ (temporal frequency), with $t$ denoting time.

We define the following auxiliary functions:
\begin{equation}
\begin{aligned}
&a(t) = \epsilon \,\sin(\omega \, t), \\
&b(t) = 1 - 2\,\epsilon \,\sin(\omega \, t), \\
&f(x,t) = a(t)\, x^2 + b(t)\, x, \\
&\frac{\partial f}{\partial x}(x,t) = 2\,a(t)\, x + b(t).
\end{aligned}
\label{eq:double-gyre-f}
\end{equation}

The time-dependent velocity field is then given by:
\begin{equation}
\begin{split}
\label{eq:double-gyre}
\mathbf{w}(x, y, t) 
&= 
\begin{bmatrix}
-\pi \, A \, \sin\bigl(\pi\,f(x,t)\bigr)\,\cos\bigl(\pi\,y\bigr)\\[8pt]
\pi \, A \, \cos\bigl(\pi\,f(x,t)\bigr)\,\sin\bigl(\pi\,y\bigr) 
\,\frac{\partial f}{\partial x}(x,t)
\end{bmatrix},
\end{split}
\end{equation}

\noindent where $A = 0.1$, $\epsilon = 0.25$, $\omega = 1$, and $t \in \mathbb{R}$. The system reduces to a steady-state Hamiltonian flow in the limit $\epsilon \rightarrow 0$.

In this benchmark, the goal is to compute the minimum-time trajectory from $\mathbf{x}_A = (1.5, 0.5)$ to $\mathbf{x}_B = (0.5, 0.5)$, assuming a unit vessel speed. Due to the time-varying nature of the flow, this scenario poses challenges for routing algorithms that must account for both spatial structure and temporal evolution.

This case directly addresses \ref{cr:dynamic-conditions}, as the optimizer must account for changing flow fields over time. Unlike static fields where all forces are known a priori, time-dependent systems require careful synchronization between spatial routing and temporal evolution. \name{} handles this by propagating time forward through the route and adjusting intermediate waypoints accordingly.

\subsection{Techy Vector Field}

This benchmark is derived from a time-dependent vector field originally proposed by Techy \textit{et al.}~\cite{Techy2011} and later adapted in maritime routing contexts by Mannarini \textit{et al.}~\cite{mannarini2019visir}. It introduces temporal dynamics into the planning problem via the following expression:

\begin{equation}
\label{eq:techy}
\mathbf{w}(x, y, t) =
\begin{bmatrix}
s x - (t - 0.5) y \\
(t - 0.5) x + s y
\end{bmatrix},
\end{equation}

\noindent where the scalar $s = -0.3$ controls the strength of a background spiral flow. The routing task consists of finding the minimal-time trajectory from $\mathbf{x}_A = \left( \cos( \pi/6 ), \sin( \pi/6 ) \right)$ to $\mathbf{x}_B = (0, 1)$, assuming a constant ship velocity of 1.

The temporal component of the vector field $\mathbf{w}(\mathbf{x}, t)$ induces a dynamic reorientation of flow, causing directional variations in the current as time progresses. The system transitions from clockwise to counter-clockwise rotation around the origin depending on whether $t < 0.5$ or $t > 0.5$. As a result, optimal routing must carefully synchronize vessel movement with these transient flow regimes to exploit favourable currents and avoid resistance.

This scenario serves as another focused test of \ref{cr:dynamic-conditions}, requiring the optimizer to respond to temporally evolving forces. Unlike purely spatial flows, the best path here depends not only on position but also on when each segment is traversed.

\subsection{Swirlys Vector Field}

Originally described by Ferraro \textit{et al.}~\cite{Ferraro2021}, the Swirlys field couples two non-linear shear components to generate a spatially varying swirl. Its mathematical definition is as follows:

\begin{equation}
\label{eq:swirlys}
\mathbf{w}(x, y) =
\begin{bmatrix}
\cos(2x - y - 6) \\
\frac{2}{3} \sin(y) + x - 3
\end{bmatrix}.
\end{equation}

The task is to compute a trajectory from $\mathbf{x}_A = (0, 0)$ to $\mathbf{x}_B = (6, 5)$ that minimizes an energy-based cost functional, subject to a fixed travel duration of $T = 30$. The cost of moving between points $\mathbf{x}_{i-1}$ and $\mathbf{x}_i$ is given by:

\begin{equation}
\label{eq:fuel-cost}
F(\mathbf{x}_{i-1}, \mathbf{x}_{i}) = \frac{1}{2} \left \| \frac{\mathbf{x}_{i} - \mathbf{x}_{i-1}}{t_{i} - t_{i-1}} - \mathbf{w}(\mathbf{x}_{i-1}, t_{i-1}) \right \|^2
\end{equation}

In the equation above, $\mathbf{w}(\mathbf{x}, t)$ denotes the current field. Although $\mathbf{w}$ is time-independent in this case, the optimizer must determine both spatial locations and the timing of each segment within the total allowed duration.

This scenario is particularly suited to evaluating \ref{cr:jit}. Due to the generous total travel time ($T = 30$), the optimizer is not simply minimizing distance, but must instead plan a route that meets the arrival deadline while minimizing energy expenditure. This encourages solutions that slow down, take detours, or even incorporate loops to remain within the timing constraint. As shown in Ferraro \textit{et al.}~\cite{Ferraro2022}, such behaviour (though rare in operational contexts) highlights an algorithm's capacity to handle Just-in-Time (JIT) constraints realistically and efficiently.

In addition, the use of a fuel-based cost function introduces a more complex objective compared to traditional time minimization. This demonstrates the flexibility of our method in handling diverse performance criteria, supporting \ref{cr:flexible-cost}. \name{}'s ability to adapt its optimization strategy to different cost structures (while simultaneously satisfying strict timing requirements) makes Swirlys a valuable test case in evaluating real-world applications.

\section{Canonical Cost Formulations}
\label{sec:appendix-cost}

For completeness, we include the two canonical benchmark objectives used in prior weather-routing studies.

\subsection{Travel-Time Objective (Constant Engine Load)}
\label{sec:cost-function-time}

Following the classical Zermelo setting~\cite{Zermelo1931}, total travel time is minimized under constant vessel speed through water $S$:

\begin{equation}
\label{eq:cost-time}
\sum_{n=1}^{L-1} \Delta t_{n} = \sum_{n=1}^{L-1} \frac{\sqrt{(x_{n+1} - x_{n})^2 + (y_{n+1} - y_{n})^2}}{w_n \cos(\psi_n) + \sqrt{S^2 - w_n^2 \sin^2(\psi_n)}}.
\end{equation}

Here $w_n$ is the local field magnitude at $(x_n, y_n, t_n)$ and

\begin{equation}
\psi_n = \arctan\left( \frac{y_{n+1} - y_{n}}{x_{n+1} - x_{n}}\right) - \arctan\left( \frac{v_{n}}{u_{n}}\right).
\end{equation}

In time-dependent fields, $t_n$ must be propagated sequentially along the route, which couples geometry and timing.

\subsection{Fuel/Energy Objective (Fixed Arrival Time)}

For Just-in-Time scenarios, total duration $T$ is fixed and we map $t_n = T r_n$ along the normalized curve parameter. Segment cost follows Ferraro \textit{et al.}~\cite{Ferraro2021}:

\begin{equation}
\label{eq:cost-fuel}
\sum_{n=1}^{L-1} F(\mathbf{x}_{n}, \mathbf{x}_{n+1}) = \sum_{n=1}^{L-1} \frac{1}{2} \left \| \frac{\mathbf{x}_{n+1} - \mathbf{x}_{n}}{t_{n+1} - t_{n}} - \mathbf{w}(\mathbf{x}_{n}, t_{n}) \right \|^2.
\end{equation}

\subsection{Feasibility Condition and Modular Objectives}
\label{sec:cost-function-other}

A basic feasibility condition is

\begin{equation}
\label{eq:speed-feasibility}
\frac{w_n}{S} < 1, \qquad \forall n \in [1, L-1].
\end{equation}

The inequality is strict: at $w_n = S$ the denominator of Equation~\eqref{eq:cost-time} can vanish for an adverse heading, so the segment time is unbounded and the vessel cannot make way against the flow.

Beyond these canonical forms, the CMA-ES stage can accept black-box objectives, including compiled or otherwise opaque vessel-performance models, because it depends only on cost evaluations. The FMS stage additionally requires first- and second-derivative information, obtained analytically, by automatic differentiation, or by a suitable numerical approximation; an opaque model is therefore compatible with the full pipeline only when those derivatives can be supplied or approximated reliably.

\section{Synthetic Land-Generation Procedure}
\label{sec:appendix-land-generation}

To stress-test obstacle avoidance, synthetic coastlines are generated with Perlin noise~\cite{perlin1985image,lagae2010survey}. A normalized scalar field over the spatial domain is thresholded by a water-level parameter to obtain a binary land-water mask. Complexity is controlled by map resolution and threshold, and reproducibility by random seed. During optimization, land interaction is evaluated by interpolating each candidate trajectory on a denser sampling and querying the mask with bilinear interpolation (implemented with \texttt{jax.scipy.ndimage.}\texttt{map\_coordinates}).

\begin{figure}[t]
    \centering
    \includegraphics[width=0.5\linewidth]{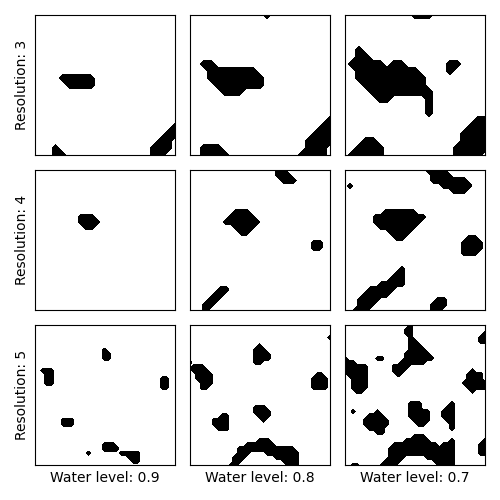}
    \caption{Different land configurations generated with Perlin noise, fixing the random seed at 4.}
    \label{fig:land-generation}
\end{figure}

Figure~\ref{fig:land-generation} shows representative configurations across increasing resolution and decreasing water level.

\subsection{Land Penalization Functional}
\label{sec:appendix-land-penalty}

This subsection gives the construction of the functional $\mathcal{P}_{\mathrm{land}}(\Gamma)$ used in Equation~\eqref{eq:land-cost}. We discretize the B\'ezier curve into a denser sequence of waypoints $\{ \mathbf{x}'_n \}_{n=1}^{L'}$ with $L' > L$, ensuring finer detection of land intersections. We define $\phi(x, y)$ as a normalized noise scalar field and $\tau \in (0, 1)$ as a water-level threshold, typically $\tau = 0.7$. Each candidate curve $\Gamma$ is interpolated across segments using $L'/L$ subdivisions per segment. Let $\mathbf{x}_{ij}$ denote the $j$-th interpolated point of segment $i$, and define the land indicator function
\[
\chi(\mathbf{x}_{ij}) =
\begin{cases}
1, & \text{if } \phi(x, y) > \tau, \\
0, & \text{otherwise.}
\end{cases}
\]

The penalization term is then
\begin{equation}
\label{eq:penalization}
\mathcal{P}_{\mathrm{land}}(\Gamma) = \sum_{i=1}^{L'-1} \max_j \chi(\mathbf{x}_{ij}),
\end{equation}
i.e.\ each segment that intersects land at any of its sub-points contributes one unit. For the real-ocean experiments, $\phi$ and $\tau$ are replaced by a land--sea mask derived from the corridor bathymetry, queried by the same dense intra-segment sampling.

\section{Vessel Performance Model}
\label{sec:appendix-performance-model}

This appendix provides an analytical reconstruction of the performance model used in the SWOPP3 benchmark (Section~\ref{sec:results-real}). The model computes instantaneous propulsive power (kW) for an 88~m cargo vessel as a function of ship speed through water $v$~(m/s), true wind speed TWS~(m/s), true wind angle TWA~(degrees, $0^\circ$ = headwind), significant wave height SWH~(m), and mean wave angle MWA~(degrees relative to heading). The formulas and constants below reproduce the compiled Rust evaluator to the error quantified in the Validation subsection.

\subsection{Validated Input Domain}

The model is calibrated and validated over the input ranges of Table~\ref{tab:perf-domain}. Evaluations outside these ranges constitute extrapolation; Section~\ref{sec:rw-setup} verifies that no segment of any reported route leaves this domain.

\begin{table}[h]
\centering
\caption{Validated input domain of the performance model.}
\label{tab:perf-domain}
\begin{tabular}{llll}
\toprule
\textbf{Input} & \textbf{Symbol} & \textbf{Unit} & \textbf{Range} \\
\midrule
True wind speed          & TWS & m/s & $[0, 30]$ \\
True wind angle          & TWA & deg & $[0, 180]$ (symmetric) \\
Significant wave height  & SWH & m   & $[0, 10]$ \\
Mean wave angle          & MWA & deg & $[0, 180]$ (symmetric) \\
Ship speed through water & $v$ & m/s & $[0, 14.5]$ \\
\bottomrule
\end{tabular}
\end{table}

\subsection{Constants}

\begin{table}[h]
\centering
\caption{Performance model constants.}
\label{tab:perf-constants}
\begin{tabular}{lll}
\toprule
\textbf{Symbol} & \textbf{Value} & \textbf{Exact fraction} \\
\midrule
$K_H$ & 4.2876\ldots & $969 / 226$ \\
$K_A$ & 0.153125 & $49 / 320$ \\
$A_W$ & 11.1395 & --- \\
$K_W$ & 0.28935\ldots & $125 / 432$ \\
$K_S$ & 0.85903125 & $27489 / 32000$ \\
Dead zone & $10^\circ$ & --- \\
\bottomrule
\end{tabular}
\end{table}

\subsection{Power Components}

Hull resistance is represented by the cubic term
\begin{equation}
\label{eq:hull}
P_{\mathrm{hull}} = K_H \cdot v^3.
\end{equation}

The apparent wind components are computed from the true wind and vessel speed as
\begin{align}
u_x &= \mathrm{TWS} \cdot \cos(\mathrm{TWA}) + v, \label{eq:ux}\\
u_y &= \mathrm{TWS} \cdot \sin(\mathrm{TWA}), \label{eq:uy}\\
V_R &= \sqrt{u_x^2 + u_y^2}, \label{eq:vr}
\end{align}
where TWA is expressed in radians for the trigonometric functions.

Aerodynamic drag contributes
\begin{equation}
\label{eq:wind}
P_{\mathrm{wind}} = K_A \cdot v \cdot \left( V_R \cdot u_x - v^2 \right).
\end{equation}

Wave added resistance is represented by
\begin{equation}
\label{eq:wave}
P_{\mathrm{wave}} = A_W \cdot \mathrm{SWH}^2 \cdot v^{3/2} \cdot \exp\!\left( -K_W \cdot |\theta_{\mathrm{MWA}}|^3 \right),
\end{equation}
where $\theta_{\mathrm{MWA}}$ is the mean wave angle in radians.

When WPS is enabled, the wingsail thrust uses the apparent wind angle
\begin{equation}
\label{eq:awa}
\mathrm{AWA} = \mathrm{atan2}\!\left( |u_y|,\; u_x \right) \quad \text{(in degrees)}.
\end{equation}
If $\mathrm{AWA} < 10^\circ$, the sail contribution is zero (dead zone). Otherwise, let $\alpha = (\mathrm{AWA} - 10^\circ) \cdot \pi / 180$, and the sail power reduction is
\begin{equation}
\label{eq:sail}
P_{\mathrm{sail}} = K_S \cdot \sin(\alpha) \cdot \left( 1 + \tfrac{3}{20}\,\sin^2(\alpha) \right) \cdot V_R^2 \cdot v.
\end{equation}

\subsection{Total Power and Energy Integration}

Without wingsails:
\begin{equation}
\label{eq:total-no-wps}
P = \max\!\left( 0,\; P_{\mathrm{hull}} + P_{\mathrm{wind}} + P_{\mathrm{wave}} \right).
\end{equation}

With wingsails (WPS):
\begin{equation}
\label{eq:total-wps}
P = \max\!\left( 0,\; P_{\mathrm{hull}} + P_{\mathrm{wind}} + P_{\mathrm{wave}} - P_{\mathrm{sail}} \right).
\end{equation}

Total voyage energy is obtained by summing over all waypoint segments:
\begin{equation}
\label{eq:energy-total}
E_{\mathrm{MWh}} = \frac{1}{1000} \sum_{i=1}^{n} P_i \cdot \Delta t_{h,i},
\end{equation}
where $\Delta t_{h,i}$ is the duration of segment $i$ in hours and $P_i$ is the power (kW) evaluated at the midpoint of each segment.

\subsection{Regularity}
\label{sec:appendix-performance-regularity}

The model is smooth in $v$ except at three loci: the clamp of Equations~\eqref{eq:total-no-wps}--\eqref{eq:total-wps}, which is non-differentiable where the pre-clamp power vanishes; the $10^\circ$ wingsail dead zone of Equation~\eqref{eq:sail}; and the absolute value in Equation~\eqref{eq:awa}, which is non-differentiable when the apparent wind is exactly fore-aft aligned. The first two arise only in WPS mode or under strong following winds; the third is specific to WPS mode, since AWA enters only through $P_{\mathrm{sail}}$. Away from these three loci the model is $C^\infty$ in $v$.

\subsection{Validation}

The analytical formulas above were validated against the compiled Rust library across the stated input domain using randomized stress testing. The maximum absolute error was 0.103~kW (mean 0.009~kW, maximum relative error 0.031\%). These results support the reconstruction as a close surrogate for independent verification, but not as a bit-for-bit reproduction of the original evaluator.

\end{document}